%% file: main.tex
\documentclass[12pt]{article}

\usepackage{amsmath,amssymb}

\usepackage{changepage}

\usepackage[utf8x]{inputenc}

\usepackage{textcomp,marvosym}

\usepackage{cite}
\usepackage{longtable}

\usepackage{nameref,hyperref}

\usepackage[right]{lineno}

\usepackage{microtype}
\DisableLigatures[f]{encoding = *, family = * }

\usepackage[table]{xcolor}

\usepackage{array}
\usepackage{xurl}
\usepackage{breakurl}
\usepackage{tablefootnote}

\newcolumntype{+}{!{\vrule width 2pt}}

\newlength\savedwidth

\usepackage{pgf,pgfplots,tikz,subfig}
\pgfplotsset{ 
  compat=newest, 
  label style = {font=\small\sffamily},
every tick label/.append style={font=\small}
  }

\usepackage[aboveskip=1pt,labelfont=bf,labelsep=period,justification=raggedright,singlelinecheck=off]{caption}

\makeatletter
\renewcommand{\@biblabel}[1]{\quad#1.}
\makeatother

\newcommand{\house}{\textrm{Hh}}
\newcommand{\work}{\textrm{W}}
\newcommand{\school}{\textrm{Sc}}
\newcommand{\retire}{\textrm{Rh}}
\newcommand{\hsp}{\textrm{Hsp}}
\newcommand{\transit}{\textrm{Tr}}
\newcommand{\leisure}{\textrm{N}}

\usepackage{lastpage,fancyhdr,graphicx}
\usepackage{epstopdf}
\begin{document}

\title{Predicting the effects of waning vaccine immunity against COVID-19 through high-resolution agent-based modeling}

\maketitle

\author{Agnieszka Truszkowska$^{1,2}$, Lorenzo Zino$^3$, Sachit Butail$^4$, Emanuele Caroppo$^{5,6}$, Zhong-Ping Jiang$^7$, Alessandro Rizzo$^{8,9}$, and Maurizio Porfiri$^{1,2,10}$}

\date{
\normalsize\vspace{.5cm} 

\noindent\textbf{1} Center for Urban Science and Progress, Tandon School of Engineering, New York University, Brooklyn NY, USA \\
\textbf{2} Department of Mechanical and Aerospace Engineering, Tandon School of Engineering, New York University,  Brooklyn NY, USA\\
\textbf{3} Faculty of Science and Engineering, University of Groningen, Groningen, The Netherlands\\
\textbf{4} Department of Mechanical Engineering, Northern Illinois University, DeKalb IL, USA\\
\textbf{5} Department of Mental Health, Local Health Unit ROMA 2, Rome, Italy\\
\textbf{6} University Research Center He.R.A., Università Cattolica del Sacro Cuore, Rome, Italy\\
\textbf{7} Department of Electrical and Computer Engineering, Tandon School of Engineering, New York University, Brooklyn NY, USA\\
\textbf{8} Department of Electronics and Telecommunications, Politecnico di Torino, Turin, Italy\\
\textbf{9} Institute for Invention, Innovation and Entrepreneurship, Tandon School of Engineering, New York University, Brooklyn NY, USA\\
\textbf{10} Department of Biomedical Engineering, Tandon School of Engineering, New York University, Brooklyn NY, USA
\\

\noindent Correspondence should be addressed to: \url{mporfiri@nyu.edu}}

\begin{abstract}

The potential waning of the vaccination immunity to COVID-19 could pose threats to public health, as it is tenable that the timing of such waning would synchronize with the near-complete restoration of normalcy. Should also testing be relaxed, we might witness a resurgent COVID-19 wave in winter 2021/2022. In response to this risk, an additional vaccine dose, the booster shot, is being administered worldwide. In a projected study with an outlook of six months, we explore the interplay between the rate at which boosters are distributed and the extent to which testing practices are implemented, using a highly granular agent-based model tuned on a medium-sized U.S. town. Theoretical projections indicate that the administration of boosters at the rate at which the vaccine is currently administered could yield a severe resurgence of the pandemic. Projections suggest that the peak levels of mid spring 2021 in the vaccination rate may prevent such a scenario to occur, although exact agreement between observations and projections should not be expected due to continuously evolving nature of the pandemics. Our study highlights the importance of testing, especially to detect asymptomatic individuals in the near future, as the release of the booster reaches full speed.

\end{abstract}

\section{Introduction}

Winter and spring 2021 marked a long-awaited massive vaccination campaign against COVID-19, starting approximately one year after the inception of the outbreak. As of the mid-September 2021, 42.6\% of the World and 63.8\% of the U.S. population took at least one dose of the vaccine, while 30.8\% and 54.5\%, respectively, are  fully vaccinated{{\cite{Mathieu2021database}}. However, approaching fall 2021 brings to light a new unknown: the possibility of waning vaccination immunity and the consequent need for an additional vaccine dose ---the booster shot{{\cite{cdc2021booster}}. There is evidence that the booster shot would not only restore the original protection, but would also enhance people's immunity against the most recent variants, including the widely dominant and highly transmittable Delta variant{{\cite{mahase2021covid,layton2021understanding}}. Many countries, including the U.S., are starting their re-vaccination campaigns, in an attempt to prevent new outbreaks accompanied by socially and economically disastrous restrictions{{\cite{mahase2021covid,mahase2021covidUK,mahase2021covidAutumn}}.

In the original (August 2021) schedule of booster shot administration by the U.S. Centers for Disease Control and Prevention (CDC), the booster campaign was expected to start on September 20\textsuperscript{th}, 2021, with booster shots available to all the adults in the U.S. eight months after they took their second vaccine dose, with plans for expansion to people taking the one-dose Johnson\&Johnson vaccine{{\cite{cdc2021booster}}. At the same time, despite a surge in new infection cases{{\cite{cdc2021cases}} and the nationwide dominance of the Delta variant{{\cite{cdc2021delta}}, non-pharmaceutical interventions (NPIs) are gradually being relaxed{{\cite{cdc2021NPIs}}, and preparations for a return to full-time in-person  education and work are underway{{\cite{usde2021schools, work2021reopen}}. Following mass vaccinations, COVID-19 testing is continuously reduced{{\cite{Hasell2020testing,cdc2021higheredtesting}}, with the enforcement of mandatory testing slowly abandoned by public health authorities{{\cite{cdc2021higheredtesting}} and contact-tracing home-isolation no longer required for fully vaccinated individuals{{\cite{cdc2021NPIs,cdc2021ctracing}}; not to mention the ongoing trend in encouraging indoor gatherings (e.g., restaurants, bars, gyms) for the fully vaccinated. In this evolving scenario, scientifically backed policy-making is of paramount importance. 

Mathematical modeling has played a key role in  assisting public health authorities to combat the COVID-19 pandemic{{\cite{estrada2020covid,vespignani2020modelling}}. Since COVID-19 onset, mathematical models are being routinely used to forecast the course of the pandemic and guide policymakers' decisions on several chief issues, including the enforcing of NPIs{{\cite{DellaRossa2020,arenas2020,Goldsztejn202opolicy,parino2021,perra2021non}}, the design of testing policies{{\cite{Du2021test,truszkowska2021covid}}, the implementation of contact tracing{{\cite{Pinotti2020,Bilinski2020,Kretzschmar2020,Quilty2021}}, and the implementation of vaccination campaigns in light of  the concurrent uplifting of NPIs{{\cite{bubar2021,Jentsch2021,Shen2021,giordano2021vaccination,Moore2021,Grundel2020,truszkowska2021designing,parino2021vaccine}}.

Mathematical modeling  can also play a critical role in the present scenario, where vaccine-induced immunity seems to be waning{{\cite{Barouch2021duration,Collier2021duration,Chemaitelly2021qatar,Goldberg2021israel}}, testing coverage is being lowered{{\cite{Hasell2020testing,cdc2021higheredtesting}}, and a booster shot campaign is going to be implemented{{\cite{cdc2021booster}}. The interplay of these critical issues has received only limited attention so far. Layton et al.{{\cite{layton2021understanding}} have simulated the emergence of new virus strains, including hypothetical deadlier variants in Ontario, Canada, in light of  realistic vaccination and booster campaigns implemented in the region. Their results, projected  until the end of 2021, point out the need of vigilance and readiness to reinstate severe NPIs, as well as the possible importance of a large-scale campaign of booster shots. Over longer time horizons, other studies have been carried out to evaluate the potential benefits of annual re-vaccination campaigns against COVID-19. In particular, Song et al.{{\cite{song2021vaccination}} have simulated different scenarios in the loss of immunity, spanning until 2029. Their findings indicate that an annual re-vaccination campaign could avoid future COVID-19 outbreaks if the vaccine is sufficiently efficacious and provides at least six months of protection. Sandmann et al.{{\cite{sandmann2021potential}} have compared the economic burden of introducing a regular vaccination program in the U.K. to the cost  associated with implementing social distancing measures for the next decade. Their work highlights the benefits of re-vaccination schemes, evidencing that they would allow to avoid large outbreaks and consequent restrictions. Lastly, Li et al.{{\cite{li2021switching}} have compared different re-vaccination strategies in 15 countries over the next 20 years in terms of long-term efficacy. Their findings identify a public health benefit in alternating re-vaccination between fragile older strata and highly active portions of the population, who habitually generate a high number of contacts. 

Although conclusive evidence on the waning immunity of the vaccine and on its timing is yet to be established{{\cite{Barouch2021duration,Collier2021duration,Chemaitelly2021qatar,Goldberg2021israel,Scott2021wane}}, these studies offer an improved understanding of the potential benefits of re-vaccination campaigns for a range of possible waning profiles. Yet, this knowledge does not immediately translate into predictions on the short-term roll-out of booster shots, which could be critical in shaping the future of the pandemic. Moreover, the long-term predictions of most of these studies are limited to coarse-grained considerations, which  cannot take into account granular details of the population.

Here, we fill in this gap by providing a systematic study of the effectiveness of a re-vaccination campaign in the ongoing 2021--2022 fall/winter season, considering as key factors the rate of administration of  booster shots and  the population coverage of testing policies implemented during this phase. We perform our study by means of a high-resolution agent-based model (ABM),  which  faithfully provides a one-to-one digital reproduction of a real, medium-sized U.S. town. As a test case, we simulate COVID-19 spreading in the town of New Rochelle, NY, for the next six months, expanding on our previous efforts published in previous issues of this journal{{\cite{truszkowska2021covid,truszkowska2021designing}}. The town of New Rochelle is chosen as a representative medium-sized US town, characterized by high levels of diversity and inequality{{\cite{USSmallTowns,dataUSA}}. The digital town closely mirrors the geography and demographics of the actual one, including  household distribution, lifestyles, and mobility patterns of its residents, thereby incorporating the diversity of its population and potential inequalities across its fabric. The progression model is expanded to include salient features of the predominant Delta variant{{\cite{cdc2021delta}}, booster shot campaign, and co-existence of three  vaccines (Johnson\&Johnson, Pfizer, and Moderna) providing  different levels of protection over time, with a gradual waning immunity. The level of detail in the model allows us to closely study the combined effect of  booster shot administration  and testing practices in this stage of the pandemic. 

The study was designed based on information about the pandemic gathered during summer 2021; some of the original design assumptions have changed during the first part of fall 2021{{\cite{cdc2021booster}}. Similar, several other changes have occurred during fall/winter 2021, including dynamical changes in the intervention policies adopted and the appearance of new variants. These changes prevent our model from faithfully reproduce the epidemic patterns observed in fall/winter 2021. However, additional simulation studies to show robustness of our qualitative findings with respect to many changes in model parameters are included as part of the Appendix to support our qualitative conclusions on the role of booster campaigns in mitigating potential COVID-19 outbreaks.

\section*{Materials and Methods}

\section{Computational framework}

Our computational framework consists of two components: a detailed database of the town of New Rochelle, NY, and a high-resolution ABM that reproduces the spread of COVID-19 at a one-to-one granularity level that includes  mobility patterns among households,  schools, workplaces, and non-essential locations (including leisure locations).

The database of the town contains geographical coordinates of every building, residential and public. Public buildings include governmental institutions and private companies of any kind, open to the general population ---the public. The database includes any workplace and non-essential locations,  identified using SafeGraph{{\cite{safegraph}}, explicitly distinguishing schools, retirement homes, and hospitals.  Town population is recreated using U.S. Census data on residents age, household and family structure, education, and employment characteristics. Residents can work and gather in New Rochelle, and in its vicinity, including New York City. They commute to work via common means such as public transit, cars, or carpools, and visit each other in private. 

Each resident of New Rochelle is mapped into an agent in the ABM, resulting in 79,205 agents. In the ABM, agents are characterized by a health state that can change according to a disease progression model detailed in the following, and they can take two types of tests --- safe, contact-less car tests, and more risky ones performed in a hospital. If infected, agents may undergo three types of treatment --- home isolation, routine hospitalization, and hospitalization in intensive care unit (ICU). The ABM was originally proposed in Truszkowska et al.{{\cite{truszkowska2021covid}}, while a later extension of the work incorporated a simplified version of the vaccination campaign{{\cite{truszkowska2021designing}}. Details about the generation of the synthetic population can be found in Section 2 of Truszkowska et al.{{\cite{truszkowska2021covid}}.

For this projective study, we tailored the ABM to capture the scenario as of fall 2021, thereby introducing realistic and time-dependent vaccination effects, booster shots, increased mobility of fully vaccinated agents, and CDC-compliant contact-tracing measures{{\cite{cdc2021ctracing,NYSCT,NRCT}}. In the following, we detail these new features. For details on the other features of the model, the reader should refer to our previous publications{{\cite{truszkowska2021covid,truszkowska2021designing}}. Figure~\ref{fig:abm} schematically illustrates major components of our computational framework.

\begin{figure}
 \centering
   \includegraphics[width=.8\textwidth]{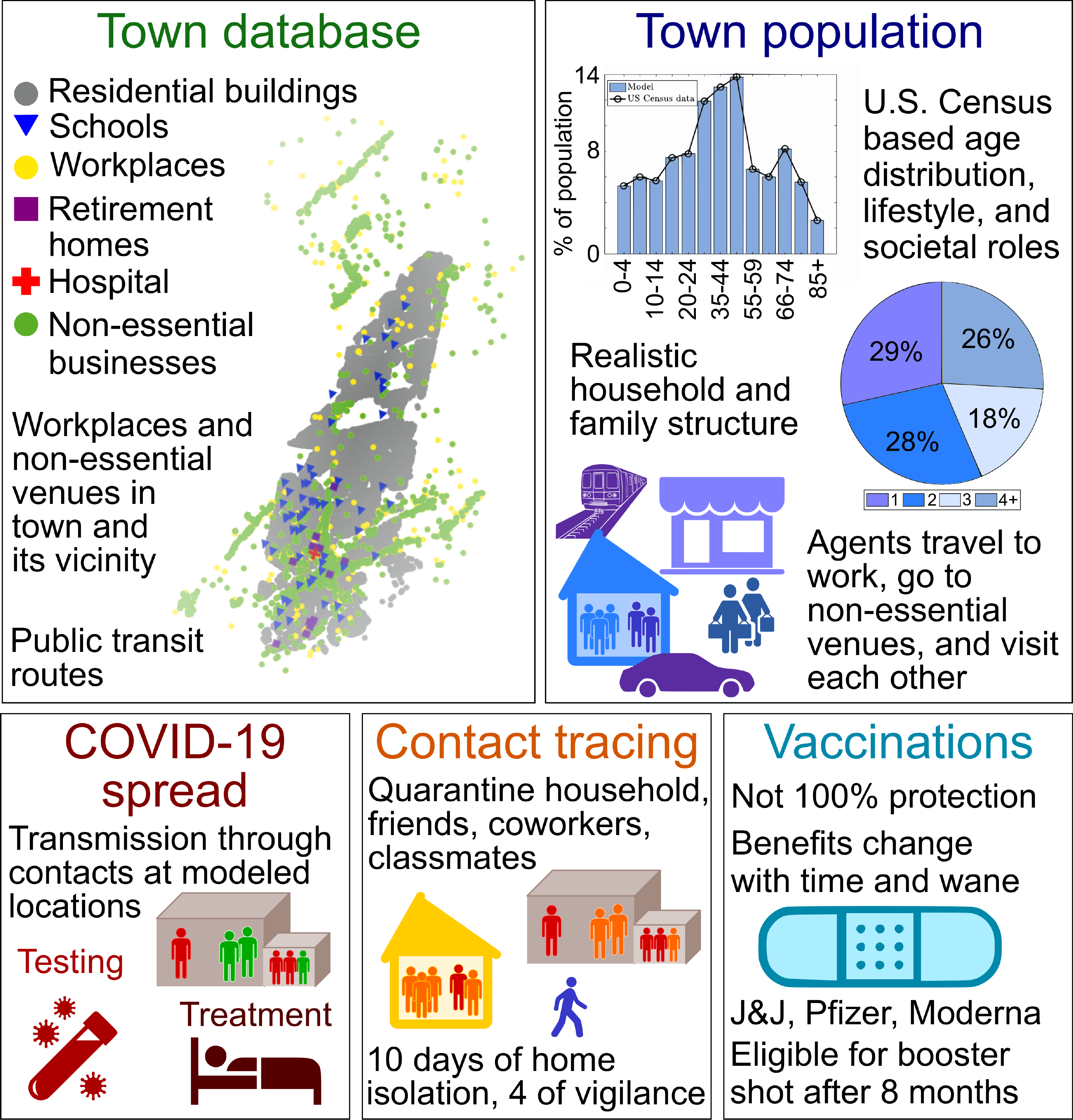}
   \caption{{\bf Schematic outline of the ABM computational framework.} The database of New Rochelle, NY, includes geographical information of every residential and public building in the town. It also incorporates workplaces and non-essential venues in the area as many town residents work outside of town and some frequent non-essential locations locations in its vicinity. Each  resident is represented as an agent. The population faithfully mirrors the sociodemographic profile of the actual one. The top-right panel shows the age distribution of agents,  as registered in the U.S. Census data. The pie chart represents  the percentage of households with the indicated size, also in close agreement with the Census (values omitted for clarity). COVID-19 spreads through contacts at different locations associated with the agents, and infected agents can be tested and treated. Positive test result triggers contact tracing, resulting in CDC-compliant home-isolation of potentially exposed individuals. Finally, the platform models imperfect, realistic vaccines, which grant a number of benefits, and wane with time. After eight months, vaccinated agents become eligible for an additional vaccine dose, the booster shot.}
   \label{fig:abm}
\end{figure}

\subsection*{COVID-19 progression model}

In our model, all the agents who are not infected, with exception of those recently recovered, are susceptible to COVID-19. Once infected,  agents can undergo testing and treatment. Agents who are not symptomatic can get vaccinated, and anyone can be contact-traced and home-isolated. 

The  progression model is shown in Fig~\ref{fig:prog}. A susceptible agent ($S$) can be vaccinated ($S_v$), may be home isolated, irrespective of their vaccination status, as a result of a home-isolation order due to a contact with an agent with a confirmed COVID-19 infection ($I_{CT}$). Isolation may also be triggered if a susceptible agent has COVID-19-like symptoms due to some other disease, such as seasonal influenza ($I_{Hm}$).  Agents can be tested, via one of the two available testing types, in a car ($T_c$) or in a hospital ($T_{Hs}$). The former type is considered contact-less and safe, while the latter carries infection risks. Complete details on testing procedures and the corresponding parameters are outlined in our two previous works{{\cite{truszkowska2021covid,truszkowska2021designing}}. Specifically, we refer to Section 3.3 of Truszkowska et al.{{\cite{truszkowska2021covid}} for more details about the testing procedures and to Table~S4 in the Supporting Information of Truszkowska et al.{{\cite{truszkowska2021designing}} for updated parameter sets. 

Susceptible individuals may become infected upon interactions with infectious individuals who are in the same building. The same building may have a role in multiple spreading pathways; for instance, a school provides pathways of infection between students, and students and teachers, but it is also the workplace for its teachers. Infections occur according to a probabilistic mechanism that accounts for differences in infection probability with respect to the characteristics of the location and the number, role, and symptomatic state of infectious individuals in the location, as detailed in Truszkowska et al.{{\cite{truszkowska2021covid,truszkowska2021designing}} (see the Appendix for more details and references). Specifically, following Ferguson et al.{{\cite{ferguson2020impact}}, we assumed that symptomatic individuals are twice as much likely to transmit the disease than asymptomatic and pre-symptomatic individuals. For non-essential locations, like leisure ones, we neglect spreading between employees and visitors, while retaining spreading within the two groups. This choice was motivated by the enforced use of personal protective equipment and social distancing toward minimizing contagions between employees and customers.

Upon infection, a susceptible agent becomes exposed ($E$), not showing symptoms of the disease. The exposed agent can also get vaccinated ($E_v$) as long as their infection status is not known. Even without any symptom, exposed agents can be tested and home isolated. Agents can either recover after being asymptomatic ($R$), or develop symptoms after the latency period and transition to the  symptomatic state ($Sy$). Symptomatic agents cannot get vaccinated, which is also the case for agents with symptoms similar to COVID-19 due to another disease. However, vaccinated agents can become symptomatic as a result of an infection ($Sy_v$), potentially leading to milder symptoms.

Agents with symptoms can test and subsequently receive treatment through home isolation ($I_{Hm}$), normal hospitalization ($H_N$), or hospitalization in an intensive care unit, ICU ($H_{ICU}$). Agents can either recover or die ($D$). Symptomatic and exposed agents can also get contact traced, and home isolated on that account. A contact-traced symptomatic agent will undergo treatment regardless of their testing status. Recovered agents are temporarily immune to COVID-19 and, after a certain period of time, they can also be vaccinated. Once their natural immunity is lost, these agents transition to the vaccinated susceptible category ($S_v$). Recovered agents who do not receive the vaccine spontaneously lose natural immunity after a fixed period of time. Based on some (possibly conservative) estimations{{\cite{Gudbjartsson2020immunity,Dan2021immunity,baraniuk2021long,Lavine2021}}, in our simulations we fixed such a period to six months. Additional simulations to assess the robustness of our findings with respect to different duration of natural immunity (loss of natural immunity after four or eight months) are reported in the Figs.~\ref{fig:4month_nat}--\ref{fig:8month_nat}.

Contact-traced agents cannot be vaccinated, and even if susceptible; they become vaccine-eligible only after some period of time. These restrictions hold for the booster shots as well. 

All the parameters that characterize the transitions in the COVID-19 progression model are listed in Tab.~\ref{tab:parameters}. An explicit expression of the contagion probability for each agent $i$, $p_i(t)$, depending on the agent's characteristics (including lifestyle, workplace or school, household in which they live) can be found in  Section 4.4 of Truszkowska et al.{{\cite{truszkowska2021designing}} The main elements of novelty of the present modeling extension include realistic treatments of the effect of vaccination and contact tracing and are detailed in the following.
\begin{figure}
 \centering
   \includegraphics[width=0.7\textwidth]{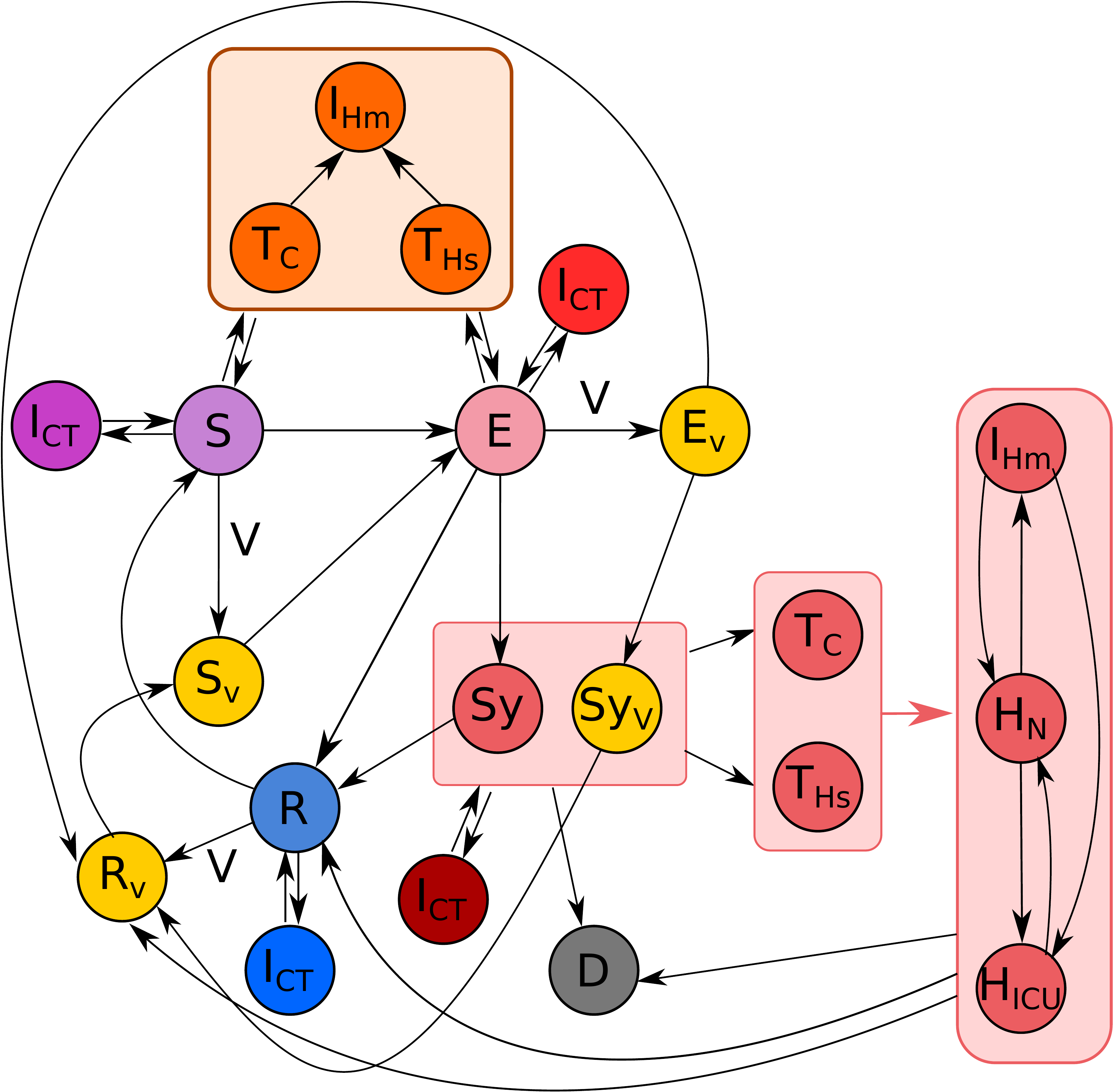}
   \caption{{\bf Diagram of the COVID-19 epidemic progression.} Agents' health states are susceptible ($S$), exposed ($E$), and symptomatic ($Sy$). Since a vaccination does not grant 100\% immunity, and exposed agents can be vaccinated, the progression distinguishes those three health states in their vaccination version, $S_v$, $E_v$, and $Sy_v$. Susceptible and exposed agents can be tested and home isolated ($I_{Hm}$). Testing can take place in a contact-less form in a car ($T_c$) or in a hospital ($T_{Hs}$). All the agents can be subject to contact tracing and subsequent home-isolation ($I_{CT}$). Exposed agent may recover without ever developing symptoms ($R$), or become symptomatic after a latency period. Symptomatic agents can undergo testing  and subsequent treatment through home isolation ($I_{Hm}$), normal hospitalization ($H_N$), or hospitalization in an intensive care unit, ICU ($H_{ICU}$). They can either recover or die ($D$). A recovered agent, if not already vaccinated, can vaccinate as well ($R_v$). Recovered agents are temporarily immune to the disease and after some period of time they become susceptible again, regardless of their vaccination status.}
   \label{fig:prog}
\end{figure}

An agent can get vaccinated with one of the three vaccine types distributed in the area according to their availability. We considered one vaccine mirroring the one-dose Johnson\&Johnson (abbreviated as $J$), and two vaccines  with  the characteristics of the two-dose Pfizer and Moderna vaccines (abbreviated as $P$ and $M$, respectively). The probability of being administered a given vaccine type was computed based on data collected manually on actual vaccine offer in the town, as of late July 2021, see Tab.~\ref{tab:vacProps}{{\cite{vacFinder}}.

Once agent $i$ is vaccinated, five of the model parameters related to the individual are modified accordingly. Specifically, four  quantities decrease upon vaccination: (1) the probability of being infected by SARS-CoV-2, (2) the transmission rate if infected, (3) the probability of requiring hospitalization, and (4) of dying if infected. Conversely,  (5) the probability of being asymptomatic when infected increases upon vaccination.

To model such a temporal effect, for each vaccine $\alpha=J,P,M$ and for each model parameter  $k=1,\dots,5$, we introduce a function $\gamma_{\alpha,k}(s)$, which models  the effect of vaccine $\alpha$ on parameter  $k$ as a multiplicative coefficient, $s$ days after vaccine administration. As an example, the probability of COVID-19 infection $p_i^v\left(t\right)$ for agent $i$  vaccinated with vaccine $\alpha$ at time $t_i$ is reduced compared to the original  probability in the absence of vaccination  $p_i\left(t\right)$ to
\begin{equation}
    p_i^v(t):= \gamma_{\alpha,1}\left(t-t_i\right)p_i\left(t\right). 
\end{equation}
Similar expressions can be written for the other four properties (see the Appendix for more details).

The shape of these functions is estimated from efficacy data on vaccines. Specifically, they are all defined as piece-wise linear functions. For the one-dose vaccine, they increase up to  their most favorable values two weeks after the shot (smaller than one for property $k=1,\dots,4$ and greater than $1$ for property $5$). In case of two-dose vaccines, the functions linearly interpolate efficacy values collected at the time of the first shot, of the second one, and at the attainment of full immunity. The second dose is always contemplated in the model, following local vaccination campaign that sets the appointment for the second shot at the time the first shot is administered, one month later{{\cite{NYSSecondDose}}. The peak benefits for all three vaccine types last for an eight-month period following recent studies on the humoral and cellular immune responses{{\cite{Collier2021duration,Barouch2021duration}}. In this period, the functions have a constant value.

The scientific community has not yet reached consensus on the duration of such period. Studies by Barouch et al.{{\cite{Barouch2021duration}} and Colliet et al.{{\cite{Collier2021duration}} provide only a lower-bound on it, whereas some preliminary analyses based on epidemic data collected over summer 2021 in countries with fast vaccination campaigns (for instance, Israel and Qatar) suggest shorter duration of peak-level immunity{{\cite{Chemaitelly2021qatar,Goldberg2021israel}}. To strengthen the robustness of our claims, parametric studies encompassing different timings of the waning vaccine immunity (six and 10 months) and a delay in the immunization effect of the vaccine are considered and discussed  in Figs.~\ref{fig:6month_s}--\ref{fig:10month} and \ref{fig:delay}. 

Once the peak-benefit period is over,  benefits start to gradually wane, yielding a gradual loss of immunity. Here, we assume that such an immunity is totally lost over the course of the following six months. This is modeled by letting the functions $\gamma_{\alpha,k}$ linearly approach $1$, over a period of six months.

Following the original CDC guidelines, we assume that people are eligible for booster shots starting from eight months after their second vaccine dose{{\cite{cdc2021booster}}. We hypothesize that the booster shot restores peak vaccination benefits immediately after its administration and beneficial effects remain constant for a period that is longer than the simulation horizon (that is, six months). The exact expressions of all the mathematical functions modeling such a phenomenon and details on their estimations are reported in the Appendix.

Agents 12 years and older can vaccinate. We model local vaccine hesitancy using an upper bound on the vaccination coverage in the town. Specifically, no more than 64,364 people are considered as eligible for vaccination (approximately the 81\% of the population), computed as a projection based on the temporal evolution of the number of new vaccinations in New York State{{\cite{Mathieu2021database,NYSTVac}}, re-scaled to the population of New Rochelle. An agent is considered fully vaccinated two weeks after their shot of a one-dose vaccine, or two weeks after the second shot of a two-dose vaccine. A fully vaccinated agent is more socially active, and is more likely to visit other agents or non-essential venues, as detailed in Tab.~\ref{tab:parameters}.

\subsection{Contact tracing}

Contact tracing is implemented in the model by complying  with local guidelines{{\cite{cdc2021ctracing,NYSCT,NRCT}}, in accordance to their stricter version issued in  winter 2021. When an agent is tested positive to COVID-19 (we contemplated a realistic quota of false positives corresponding to 5\% of the tests{{\cite{Healy2021}}), their household members and all the agents with whom they carpool, in case this is their transit mode to work, are immediately home-isolated. 

Moreover, a predetermined number of coworkers is home-isolated. To account for realistic implementation of contact tracing, we bound the maximum number of home-isolated coworkers to a given value of 10 and the same upper bound is used throughout for schools and residents. In particular, contact tracing of a retirement home employee results in home-isolating 10 residents in addition to coworkers. Conversely, a confirmed positive resident leads to home-isolating 10 other residents and employees. With respect to schools, the granularity of our model was set to the single school. Hence, contact tracing of a student who tested positive is modeled by home-isolating 10 students of the same age from that agent's school, plus one teacher. The same logic applies also upon tracing a teacher, with a random choice of 10 same-aged students to be home-isolated. 

Finally, since agents visit each other in private, we model contact tracing imposing home-isolation to the entire households visited by a COVID-19 positive agent during the course of 14 days preceding the time the agent was determined positive, according to local policies. Due to the limited supervision on restrictions to private visits, we accounted for reduced compliance, estimating such a parameter from the literature, see Tab.~\ref{tab:parameters}. 

In the model, home-isolation is implemented by placing the agent in home isolation for a period of 10 days. Afterwards, the agent continues to monitor themselves for COVID-19 symptoms for a duration of 4 days, reflecting the guidelines. If during this two-week period the agent develops COVID-19 symptoms, they are assigned to an adequate treatment, regardless their testing status. Finally, following the  guidelines, fully vaccinated agents still have to home-isolate, and negative test results do not shorten the home-isolation duration.

\subsection{Simulation setup}

Simulations are initialized with a predetermined number of COVID-19 infected agents in the two phases of the disease, that is, exposed or symptomatic, to mimic real conditions in the town. These initial cases can be in different testing stages and undergo treatment. An initial number of vaccinated agents is also contemplated, based on the data collected from the vaccination campaign put in place between January 2021 and the start of the simulation. We assume that each of the 51,342 individuals already vaccinated at the beginning of the simulations has received their first shot in a randomly chosen day between the beginning of the vaccination campaign in January 2021 and September 7\textsuperscript{th} 2021 (see the Appendix for the temporal distribution of first shots), resulting in different level of immunity at the beginning of the simulations for these vaccinated agents. In the Appendix, we present additional simulations to assess the robustness of our findings with respect to different approximations of the temporal distribution of first shots (see Fig.~\ref{fig:distribution}). 

Model parameters related to vaccinations and contact tracing are based on the literature and official releases from the CDC{{\cite{CDCCovid}}, as detailed in the above. The characteristics of different vaccine types are based on official CDC and Food and Drug Administration (FDA) releases{{\cite{cdcJJ, cdcPfizer, cdcModerna, fdaJJ,fdaPfizer,fdaModerna}} and are outlined in detail in the Appendix. As indicated therein, in the absence of confirmed values, we either interpolated between the known benefit levels, or we used them for scaling. The parameters used in our contact tracing practices are also listed in Tab.~\ref{tab:parameters}, where our assumptions on the number of contacts each agent has in their workplaces, schools, and other  visited locations, are detailed. The complete parameter set and all the modeling assumptions are detailed in Tab.~\ref{tab:parameters}

\section{Results}

Our simulations projected COVID-19 spreading over a time span of six months starting from September 7\textsuperscript{th} 2021. At this time, most of the town residents eligible for a vaccine had received their vaccination earlier in the year. Specifically, 51,342 residents were  vaccinated with at least one dose as of September 7\textsuperscript{th} 2021{{\cite{NYSTVac}}. As the first dose was administered in January 2021, during the six-month simulation window many of the vaccinated residents would lose their immunity (see Fig.~\ref{fig:day_wane}). The types of the vaccines and their effects mirrored those that were distributed in the area and included the two double-dose vaccines (Moderna and Pfizer) and one single-dose vaccine (Johnson\&Johnson), see  Tab.~\ref{tab:vacProps}. Per the original, August 2021 CDC guidelines, an agent was set to start losing their immunity at approximately eight months after they become fully vaccinated{{\cite{cdc2021booster}}. At this time, they become eligible for a booster shot, which would restore their peak resistance to the virus, thereby immunizing again the population at the rate set by the administration. Booster shots in the model are distributed alongside regular vaccination doses. In every simulation, only a fixed number of shots can be administered each day, in the form of booster or first shots, with no particular prioritization. For example, a rate of twenty vaccines per day implies that twenty randomly chosen, eligible agents will receive their vaccine dose that day, either their first or their booster shot, according to their vaccination status.

\subsection{Curbing an upcoming wave requires a vaccination rate at least equal to the rate in spring 2021}

To quantify the impact of the vaccination rate on the spread of COVID-19, we performed simulations with two different rates: 0.58\% and 0.11\%  of the total population per day. These two values correspond to the maximum first-dose vaccination rate attained at the beginning of April 2021 and the rate registered in early September 2021, respectively{{\cite{NYSTVac}}. The former represents an optimal scenario, which can be achieved only if local authorities implement large, temporary vaccination centers or other viable alternatives; the latter could be considered as a worst case scenario of low vaccination rate. 

In our simulations, whose outcome is illustrated in Fig.~\ref{fig:6month}, we assumed that highly effective testing practices were enacted during the entire period. In particular, we hypothesized that each symptomatic agent was tested with probability equal to $80\%$, while such a probability was reduced to $40\%$ for asymptomatic agents. These parameters are representative of optimal testing practices{{\cite{Pullano2021}}, and they are used to illustrate that, even under optimistic assumptions on the efficacy of testing practices, low vaccination rates may lead to tremendous increases in infections and death toll.

\begin{figure}
 \centering
   \includegraphics[width=.85\textwidth]{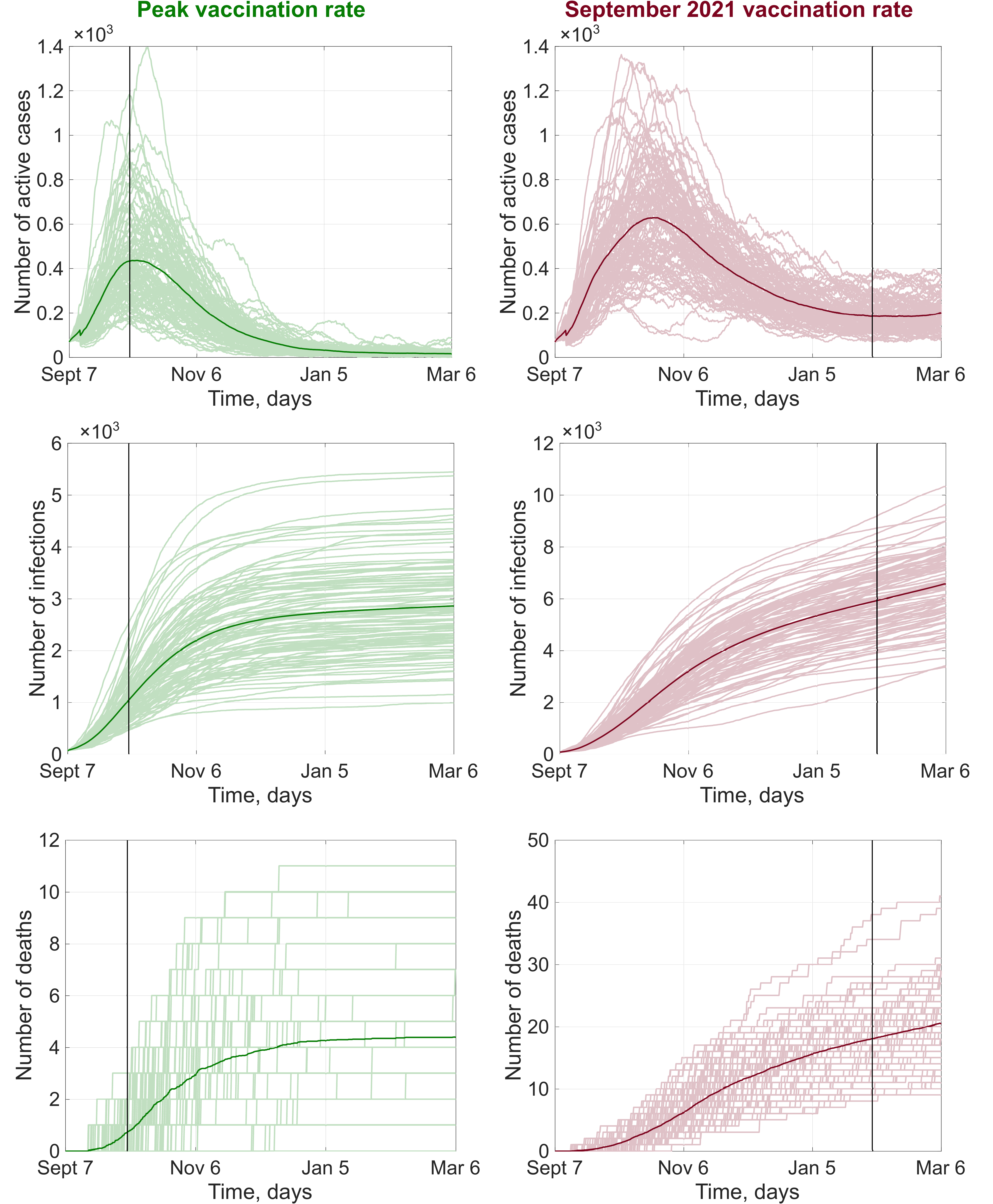}
   \caption{{\bf COVID-19 spreading over six months from September 7\textsuperscript{th} 2021, amid two different vaccination campaigns.} Active cases, total number of infections, and total deaths for the next six months at either peak vaccination rate of 0.58\% population/day (green) or present vaccination rate of 0.11\% population/day (red). For each scenario, $100$ independent realizations are shown and their average is highlighted. The vertical lines denote the date at which the entire non-hesitant eligible population is expected to be vaccinated with at least one shot. }
   \label{fig:6month}
\end{figure}

We compared the number of infections and death toll for the two vaccination rates for six months starting from September 7\textsuperscript{th}, 2021. Results from Fig.~\ref{fig:6month} show that, for the higher vaccination rate (green curves), the number of active cases should start decreasing from mid-October. The average peak of active cases should exceed 400 active cases per day, and then it should quickly drop in few weeks, potentially reaching the end of the outbreak at the beginning of 2022. On the contrary, the current vaccination rate (red curves) would lead to a 50\% increase in number of cases per day during fall 2021. Even more alarming is the projection that it would not be sufficient to eradicate the disease, leading to a possible slow rise in number of cases during winter 2022, and potentially a resurgent wave in spring 2022. These results indicate the need to maintain a fast pace during the booster campaign toward curbing potential upcoming waves and quickly eradicating the disease. 

In all the simulations, we observed an initial phase in which the number of cases steadily increases. We believe that such an increase could be caused by an underestimation of the initial number of infected individuals, due to under-detection in the officially reported data used to initialize the simulations. However, such an initial increase does not impact our insights into the effects of waning immunity, as more than 88\% of the individuals vaccinated during spring and summer 2021 has still full immunity at the end of October 2021 (see Fig.~\ref{fig:day_wane}). In support to these insights, we performed additional simulations that demonstrate robustness of our findings with respect to different assumptions on the initial number of infected individuals. The outcome of these simulations is shown in Figs.~\ref{fig:doubleI} and~\ref{fig:fiveI}.

\subsection{Testing is still needed, even with high  vaccination rates}

We also investigated the role of testing and contact tracing implemented during the booster shot campaign, toward elucidating the impact of these practices, their interplay with the vaccination rate, and, ultimately, to understand whether massive testing campaigns are still needed in this phase. 

We conducted a parametric study by varying the vaccination rate and the overall efficacy of testing practices over a two-dimensional grid. Specifically, we considered re-vaccination rates ranging between 0.01--5\% of the population per day. These two extreme values represent scenarios in which the entire re-vaccination campaign would last more than 20 years or just 20 days. For context, the first-dose peak vaccination rate was 0.58\% during April 2021 and the lowest rate was 0.027\% in mid-summer 2021{{\cite{NYSTVac}}. The efficacy of the testing practices was encapsulated by a global parameter, termed ``testing efficacy,'' which measures the probability that a symptomatic agent is  tested. In the simulations, we varied such a parameter from 10\% to 100\%, representing scattered to ideal testing. 

We performed these parametric studies within three different detection scenarios, according to the ability of detecting pre-symptomatic and asymptomatic agents (hereby, referred to as exposed): high detection (in which exposed agents are tested with the same probability of symptomatic ones), medium detection (in which the probability for an exposed individual to be tested is reduced by 50\% with respect to the one of a symptomatic agent), and low detection (in which exposed agents reduce the probability of being tested to 10\% of the one of symptomatic agents). While high detection of exposed is ideal ---but likely unrealistic, since asymptomatic infections are more difficult to be detected without a massive implementation of testing practices and contact tracing--- medium detection could be a realistic proxy of testing practices seen since the onset of the pandemic{{\cite{Pullano2021}}, and low detection could potentially represent a scenario in which most routine testing practices are disbanded. 

Our results, shown in Fig~\ref{fig:heat}, highlight the need to continue testing during the upcoming booster shot campaign. In particular, for all the examined detection scenarios, testing less than 20--30\% of symptomatic agents always resulted in a dramatic increase of infections and deaths. To overcome the ensuing surge it would necessary to apply unprecedentedly high vaccination rates of 1--5\% of the total population per day, likely beyond the capacity of the healthcare system that we have seen in spring 2021.

\begin{figure}
 \centering
   \includegraphics[width=\textwidth]{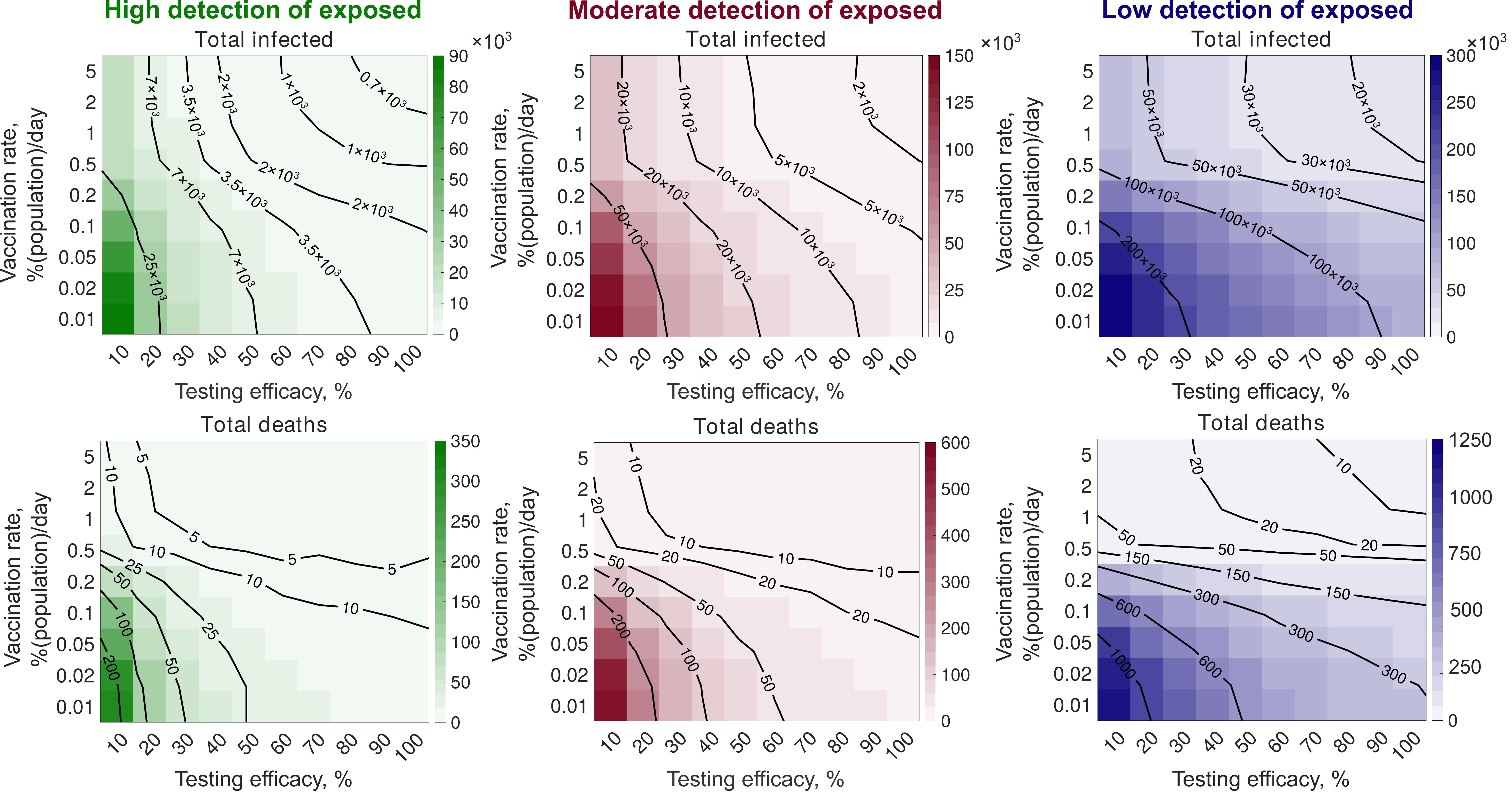}
   \caption{{\bf Interplay between re-vaccination rates and testing efficacy.} Two-dimensional heat-maps showing the combined effect of vaccination rate and testing efficacy on the total number of infected and deaths over a period of six months starting from September 7\textsuperscript{th} 2021. Three different detection levels of exposed agents capture a range of contact tracing efforts. 
   }
   \label{fig:heat}
\end{figure}

Our results also emphasize that detecting pre-symptomatic and asymptomatic agents is a critical issue. In fact, for all combinations of re-vaccination rate and testing efficacy, reduced detection of such agents results in a many-fold increase of total number of infections and deaths. For example, with low detection of exposed agents (third scenario, in blue in Fig.~\ref{fig:heat}), the number of deaths may exceed over 600  (that is, approximately 0.8\% of the population of the town), reaching peaks of more than 1,000 deaths in the worst case scenarios of both low testing efficacy and low re-vaccination rates. Further evidence on the key role of contact tracing is offered through an additional set of simulations (reported in Fig.~\ref{fig:nct}), in which no contact tracing practices are modeled. Results of these simulations suggest that in the absence of any form of contact tracing the COVID-19 death toll would dramatically increase, even in the scenario of fast re-vaccination rates.  

\section{Discussion and conclusion}

The chief goal of this work was to systematically analyze the spread of COVID-19 in the upcoming 2021 fall/winter season, as immunity gained due to vaccination wanes over the year and  testing  practices change. Toward this aim, we extended a mathematical model designed in our previous efforts{{\cite{truszkowska2021covid,truszkowska2021designing}}, a high-resolution ABM of a medium-sized U.S. town faithfully reproducing spatial layout, demographics, and lifestyles of  urban areas, to quantify the effects of a range of vaccination and testing efforts. As in our previous studies, we focused on the town of New Rochelle, NY, which was the location of one of the first COVID-19 outbreaks in the U.S.. New Rochelle is representative of many towns in the country and is characterized by high levels of diversity and potential inequalities{{\cite{USSmallTowns,dataUSA}}.

Complementing our earlier efforts, we enhanced the capabilities of the computational framework along three main directions. First, we considered realistic types and administration of vaccines, as well as time-varying vaccination benefits, including waning immunity after a tunable period{{\cite{Barouch2021duration,Collier2021duration,Chemaitelly2021qatar,Goldberg2021israel}} and administration of a booster shot{{\cite{cdc2021booster}}. Second, natural immunity achieved through recovery was also considered to be no longer permanent{{\cite{Gudbjartsson2020immunity,Dan2021immunity}}. Third, we modeled contact tracing, consistent with the CDC and local health department guidelines{{\cite{cdc2021ctracing,NYSCT,NRCT}}. Overall, the current model is a highly realistic and detailed digital representation of the town and its residents, with the resolution of a single individual, thus allowing for reliable ``what-if'' analyses of the epidemic during the upcoming fall/winter season. Equipped with a new parameter set tuned on the now-dominant Delta variant, we studied the local outcome of the interplay between the rate of vaccination and efficacy of testing practices.

Predictably, we found that low testing efficacy may lead to a disastrous increase in both infections and deaths, irrespective of vaccination efforts of any intensity. In fact, low testing efficacy seems to hamper any benefits that would be offered by realistic re-vaccination campaigns. The final count of cases and casualties would be substantially independent of vaccination rates, unless booster shots were administered to more than 1\% population per day (an unrealistic scenario, since it exceeds the peak vaccination rate during spring 2021). For low-to-moderate testing efficacy, vaccination rates below 0.5\% consistently lead to a case and death toll comparable with those experienced during the first wave{{\cite{truszkowska2021covid}}.

These results, in agreement with other studies on testing practices during previous phases of the COVID-19 pandemic{{\cite{aleta2020modeling, Quilty2021}}, highlight the central role of testing, contact tracing, and home-isolation in the fight against COVID-19 and echo the ``Path out of the Pandemic," presented by the U.S. Government on September 10\textsuperscript{th}, 2021, as part of ``President Biden's COVID-19 Plan"{{\cite{WhiteHousePlan}}.

To contain COVID-19 mortality below the level of the first wave, we predict that at least 0.5\% of population per day should be immunized/re-immunized, as testing and contact tracing are carried out with moderate efficacy. Such a 0.5\% vaccination rate is not unreasonable, as it is comparable to the average vaccination rate during the peak of the spring 2021 vaccination campaign{{\cite{NYSTVac}}. However, such a peak vaccination rate was accompanied by large, temporary vaccination centers that no longer exist. Hence, local authorities might need to restore these temporary vaccination centers or provide viable alternatives, to keep the administration of boosters at the desired rate. On the contrary, vaccination rates below 0.5\% might lead to scenarios that are worse than those recorded in spring 2020{{\cite{OWD_us}}. In particular, using a vaccination rate equal to that adopted in September 2021 would lead to a potentially disastrous rise in the number of infections around the beginning of 2022. While the number of deaths projected in this scenario are still lower than those occurred in the first wave, likely due to reduced mortality rates of vaccinated individuals, the steep increase portends that this number would ultimately overcome first wave figures. 

These projections emphasize the importance for a booster shot, in line with the ``President Biden's COVID-19 Plan"{{\cite{WhiteHousePlan}} that highlights the need of ``further protecting the vaccinated'' (with the booster shot). To efficiently combat the spread, the booster shot campaign should be conducted on a scale close to the one implemented during the peak immunization efforts in spring 2021. Similar conclusions have been drawn by other authors. For example, Layton et al.{{\cite{layton2021understanding}} report doubling of deaths by late December 2021 in Ontario, Canada, as a consequence of reducing the baseline vaccination rate by 20\%. Sandmann et al.{{\cite{sandmann2021potential}} predict the occurrence of up to two annual COVID-19 waves in the UK, whose magnitudes are strictly tied to vaccine efficacy and active NPIs. In the worst case scenario, it is expected that there will be a new wave this fall, with a magnitude comparable, or even higher, than the one observed during 2020. Similarly, Song et al.{{\cite{song2021vaccination}} indicate reoccurring new surges in the worst cases of vaccination efficacy and immunity duration, and a constant, but non-zero COVID-19 incidence in the best scenarios, starting from mid-2021. 

Testing of symptomatic individuals plays a key role in controlling the spread, especially when it is accompanied by moderate contact tracing efforts. Seen from another perspective, testing a mere 40\% of the symptomatic individuals with moderate contact tracing efforts should avoid exceeding mortality rates of the first wave. Beyond a 60\% testing efficacy, the effect of increased testing is diluted and higher vaccination rates are needed to bring` down mortality rates. While testing levels of 40\% or above are achievable{{\cite{TestsJH}}, as they are comparable with the estimates for the late summer 2020 in France{{\cite{Pullano2021}} they are still challenging to attain. Reducing delays in testing and contact tracing could offer a pathway to mitigate difficulties in reaching high testing levels{{\cite{Kretzschmar2020,Quilty2021}}.

Likewise, the detection of asymptomatic individuals is of paramount importance to combat the spreading. In particular, going from high- to low-detection of such individuals more than doubles the number of cases and deaths. This finding is consistent with the literature, whereby efficacious tracking of the asymptomatic individuals has been shown to arrest the progression of the spread of the virus{{\cite{reynalara2020virus, kretzschmar2021isolation}}. High detection rates can be realized with aggressive contact tracing strategies that can identify stranger contacts in addition to close contacts{{\cite{rodriguez2021population}}. At the same time, while it is reasonable that most people who develop symptoms or are informed of exposure to an infected individual will isolate, and possibly test, detecting asymptomatic individuals could become progressively more difficult, especially with general decline in social distancing practices and lifting of mandatory testing by many employers and institutions{{\cite{cdc2021higheredtesting}}.

While insightful, our results are not free from limitations. Though calibrated in real data, the  high granularity of our model comes at a cost of a series of assumptions. Importantly, immunity due to vaccination was modeled based on educated guesses due to limited data availability. Except for waning immunity benefits from vaccination, all the parameters in our simulations were time-invariant; in real settings factors such as NPIs or testing coverage are likely to change in response to emerging situations{{\cite{NYSClusters, NYSMasksNew}} and, likewise, vaccination rates to dynamically change. Moreover, we tested the general, uninfected population in a non-random fashion, and contact tracing guidelines within our model were more conservative than those currently in-place. Finally, in our ABM, we implemented vaccination assuming no prioritization based on age or between booster shots and first shots. The analysis in Massonnaud et al.{{\cite{Massonnaud2021.12.01.21267122}} with a compartmental model suggests that different prioritization strategies may impact the COVID-19 toll, calling for future work toward investigating this problem with our highly granular ABM. 

Concerning the timing and profile of waning immunity, in our study we made several assumptions based on the knowledge available at the time of writing the paper. We acknowledge that the scientific community has yet to reach complete consensus. Specifically, we set immunity benefits from vaccination to start to gradually wane after a period of eight months from peak-level immunity. This is in accordance with  recent studies on the humoral and cellular immune responses, which indicates eight-months as a lower-bound on this period{{\cite{Barouch2021duration,Collier2021duration}}. However, other studies suggest different, and potentially shorter, timings{{\cite{Chemaitelly2021qatar,Goldberg2021israel}}, thereby  conclusive evidence is yet to be established{{\cite{Scott2021wane}}. Similar uncertainty seems to be present on the duration of natural immunity{{\cite{Gudbjartsson2020immunity,Dan2021immunity,baraniuk2021long,Lavine2021}}, which, in this work, was chosen to last for six months. To partially address these uncertainties, we performed a parametric study that is reported in the Appendix, which ensure that our qualitative findings and observations are robust to changes in the timing and profile of the waning immunity. 

The study design was based on information about the pandemic gathered during summer 2021. In particular, in the original (August 2021) schedule, booster shots were planned to be available to all the adults in the U.S. eight months after they took their second vaccine dose{{\cite{cdc2021booster}}. This schedule has changed several times, as currently COVID-19 vaccine booster shots are available for some categories of people who completed their initial series at least six months ago (for Pfizer and Moderna), or two months ago (for Johnson\&Johnson){{\cite{cdc_booster_november}}. New changes to such a plan are expected in the near future, as the ``President Biden's COVID-19 Plan" suggests ``to quickly get booster shots into the arms of eligible Americans once approved"{{\cite{WhiteHousePlan}}. As scenarios are  rapidly changing in the U.S. and throughout the globe, we have opted to adhere to the original CDC guidelines for our simulations. We believe that the additional simulations reported in Figs.~\ref{fig:6month_s}--\ref{fig:10month} provide some insights into this issue, suggesting that the rate of vaccination is more important than its actual timing, to avoid potential, resurgent outbreaks in late winter/spring 2022.

We acknowledge that our model, in its current form, does not accurately reproduce the epidemic data observed in fall/winter 2021. At the time of our simulations and writing, we did not observe the surge of the Omicron variant, while  testing and containment policies have dynamically changed several times during the entire simulation window. However, the additional simulations reported in the Appendix confirm robustness of our findings with respect to many potential confounds, supporting the qualitative reliably of our projections. 

The need to administer booster shots must also be put in context with respect to medical, social, and moral concerns{{\cite{mahase2021covid, schaefer2021making}}. First, the waning of immunity is still not confirmed with certainty\cite{Scott2021wane}, and the health effects of an additional dose remain, to some extent, unexplored{{\cite{mahase2021covid}}. It cannot be excluded that an additional dose may only selectively boost the efficacy for individuals who are immunocompromised or whose initial vaccination had low efficacy{{\cite{krause2021considerations}}. Also, any adverse effects of the booster dose may have a negative impact to the vaccine acceptance{{\cite{krause2021considerations}}. Second, with less than 5\% of the populations in low income countries being fully vaccinated, the World Health Organization has deemed every booster shot as ``ethically questionable'' and warned that unmitigated COVID-19 pandemic in those areas will continue yielding new variants{{\cite{schaefer2021making, who2021booster}}.
Despite these concerns, countries have already started their booster shot campaigns in an attempt to curb the risk of new surges and restrictions{{\cite{wadman2021israel}}. These decisions are likely driven by the Delta variant, which dilutes the herd-immunity thresholds estimated for the wild-type strain{{\cite{randolph2020herd,kwok2020herd,kadkhoda2021herd,macintyre2021modelling}}.

\section*{Acknowledgments}

We would like to acknowledge Maya Fayed and Sihan (Silvia) Wei for updating the town database, identifying part of the new model parameters, and introducing the code for the out-of-town non-essential locations. The work of AT and MP was partially supported by National Science Foundation (CMMI-1561134 and CMMI-2027990). The work of EC, ZPJ, and AR was partially supported by National Science Foundation (CMMI-2027990). The work of SB was partially supported by National Science Foundation (CMMI-2027988). The work of AR was partially supported by Compagnia di San Paolo. The funders had no role in study design, data collection and analysis, decision to publish, or preparation of the manuscript.

\section*{Authors contribution}
Conceptualization---AT, LZ, SB, AR, MP; 
data curation---AT; 
methodology---AT, LZ, SB, AR, MP; 
software---AT, SB; 
validation---AT; 
formal analysis---all the authors;
investigation---all the authors; 
resources---MP; 
writing—original draft preparation---AT, LZ, SB, AR, MP; 
writing—review and editing---EC, ZPJ; 
visualization---AT; 
supervision---SB, EC, ZPJ, AR, MP; 
project administration---MP; 
funding acquisition---SB, ZPJ,AR, MP.


\appendix

\section{Details about the model}\label{sec:details}

Here, we detail additional information on the model, building on our previous work \cite{truszkowska2021covid,truszkowska2021designing}. In the following, we will refer to these two publications and, when possible, to exact locations within these references for details about the framework and progression. We focus the presentation on the general formulation of the COVID-19 transmission model and on the new elements introduced in this study and changes with respect to the previous implementations (such as those due to the Delta variant).

\subsection{COVID-19 transmission}\label{sec:transmission}

We start by briefly summarizing the COVID-19 transmission and progression model proposed in our previous publications \cite{truszkowska2021covid,truszkowska2021designing}. In our model, a susceptible ($S$) agent $i$ becomes infected with COVID-19 (and thus exposed, $E$) according to a probabilistic rule. Specifically, at time $t$, the agent becomes infected with probability equal to
\begin{equation}
    p_i(t):= 1 - e^{-\Delta t \Lambda_i\left(t\right)},
\label{eq:pii}
\end{equation}
where $\Delta t=0.25$ day is the duration of a discrete time-step and $\Lambda_i\left(t\right)$ is the overall infectiousness of the locations that the agent is associated with, that is, the locations that the agent may visit. The function $\Lambda_i\left(t\right)$ accounts for contributions from the different types of locations associated with agent $i$ and can be written as
\begin{equation}
\begin{aligned}
    \Lambda_i\left(t\right) := &\lambda_{\house,f_{\house}(i)}\left(t\right) + \lambda_{\work,f_{\work}(i)}\left(t\right) + \lambda_{\school,f_{\school}(i)}\left(t\right) + \lambda_{\retire,f_{\retire}(i)}\left(t\right) \\ &+ \lambda_{\hsp,f_{\hsp}(i)}\left(t\right) + \lambda_{\transit,f_{\transit}(i)}\left(t\right) + \lambda_{\leisure,f_{\leisure}(i,t)}\left(t\right),
\end{aligned}
\label{eq:lam}
\end{equation}
where the function $\lambda_{\bullet,\ell}(t)$ is the infectiousness of location $\ell$ at time $t$. The first subscript denotes the type of location, that is, households ($\house$), workplaces ($\work$), school buildings ($\school$), retirement homes ($\retire$), hospitals ($\hsp$), public transit ($\transit$), and  non-essential locations ($\leisure$); the function in the second subscript, $f_\bullet(i)$, selects the location of a given type, which agent $i$ is assigned to. Further details and the explicit expressions of the infectiousness functions in houses, workplaces, schools, retirement homes, and hospitals are reported in  Section 3.2 of~\cite{truszkowska2021covid} (with complete details in Appendix~\ref{sec:details}); details on the infectiousness of public transit and non-essential locations can be found in Section 4.4 of~\cite{truszkowska2021designing} (with complete details in Appendix~\ref{sec:details}).

\subsection{Vaccination}

In the following, we expand the corresponding subsection in the Materials and Methods with details on modeling the effects of vaccinations on COVID-19 progression. Specifically, once an agent $i$ is vaccinated with the first dose, five parameters in the original model \cite{truszkowska2021covid} related to that individual are modified: (1) the probability of being infected by COVID-19, (2) the transmission rate, (3) the probability of requiring hospitalization, (4) the probability of dying, and (5) the probability of being asymptomatic. 

The extent to which these five parameters are impacted for an agent $i$ depends on the vaccine type $\alpha\in\{J,P,M\}$ and the number of days elapsed since the vaccine was administered. Specifically, a parameter (or function) $k = \{1, 2, 3,4, 5\}$ is modified by vaccine $\alpha$, $s$ days after the first dose is administered through a function $\gamma_{\alpha,k}(s)$. 
The functions corresponding to the five parameters are detailed next.

The probability of being infected by COVID-19 for a susceptible unvaccinated agent $i$ is originally defined in \eqref{eq:pii}. Upon being vaccinated vaccine $\alpha$ at day $t_i$ this probability is reduced to
\begin{equation}
    p_i^v(t):= \gamma_{\alpha,1}\left(t-t_i\right)p_i\left(t\right).
\label{eq:pii2}
\end{equation}

Along the same lines, once an agent $i$ gets infected with COVID-19, their transmission rate becomes
\begin{equation}
    \beta_i^v(t):= \gamma_{\alpha,2}\left(t-t_i\right)\beta_i\left(t\right),
\label{eq:beta}
\end{equation}
where $\beta_i(t)$ is the transmission rate for an unvaccinated agent. The probability of requiring hospitalization is similarly reduced compared to its base value $\chi_i$ to
\begin{equation}
    \chi_i^v(t):= \gamma_{\alpha,3}\left(t-t_i\right)\chi_i,
\label{eq:hsp}
\end{equation}
and the probability of dying decreases from that for an unvaccinated agent at $\mu_i$ to
\begin{equation}
    \mu_i^v(t):= \gamma_{\alpha,4}\left(t-t_i\right)\mu_i.
\label{eq:mrt}
\end{equation}
Finally, the probability of becoming asymptomatic for a vaccinated agent increases from $\sigma_i$ according to
\begin{equation}
    \sigma_i^v(t):= \gamma_{\alpha,5}\left(t-t_i\right)\sigma_i.
\label{eq:asy}
\end{equation}

For unvaccinated agents, the probabilities of hospitalization $\chi_i$, dying $\mu_i$, and becoming asymptomatic $\sigma_i$ depend only on testing practices and age, and are therefore independent of $t$. Instead they depend on time for vaccinated agents, as illustrated in Eqs.~\eqref{eq:hsp}--\eqref{eq:asy}.

The functions $\gamma_{\alpha,k}(s)$ have a piece-wise linear form, controlled by $k$ and $\alpha$. Specifically, the functions are designed to reach a peak value in $14$ days after the single shot of Johnson\&Johnson, and in $44$ days for the two-dose vaccines. Functions decrease for parameters $k=1,\dots,4$ and increase for parameter $k=5$. Since, the peak benefits from vaccines last for an eight-month period (following CDC recommendations for when a booster shot should be taken\cite{cdc2021booster}), the functions are designed to attain a constant value in this window. Once the corresponding $254$ day period after the vaccination is over, the functions linearly interpolate to 1 over the course of six months (that is, until day $434$), beyond which they remain at 1. Hence, the curve is fully determined by two parameters: the value of the function immediately following vaccination ($\Gamma_0$) and the peak value ($\Gamma_{14}$), as illustrated in Fig.~\ref{fig:1dose}. These values are reported in Table~\ref{tab:vacJJ}.

\begin{figure}
    \centering
\subfloat[]{
   \input{1dose}\label{fig:1dose}}
  \\
  \subfloat[]{
   \input{2dose}\label{fig:2dose}}
    \caption{Shape of the functions $\gamma$ for {\bf (a)} a one-dose vaccine (in blue) and for {\bf (b)} a two-dose vaccine (in green). }
    \label{fig:k3}
\end{figure}
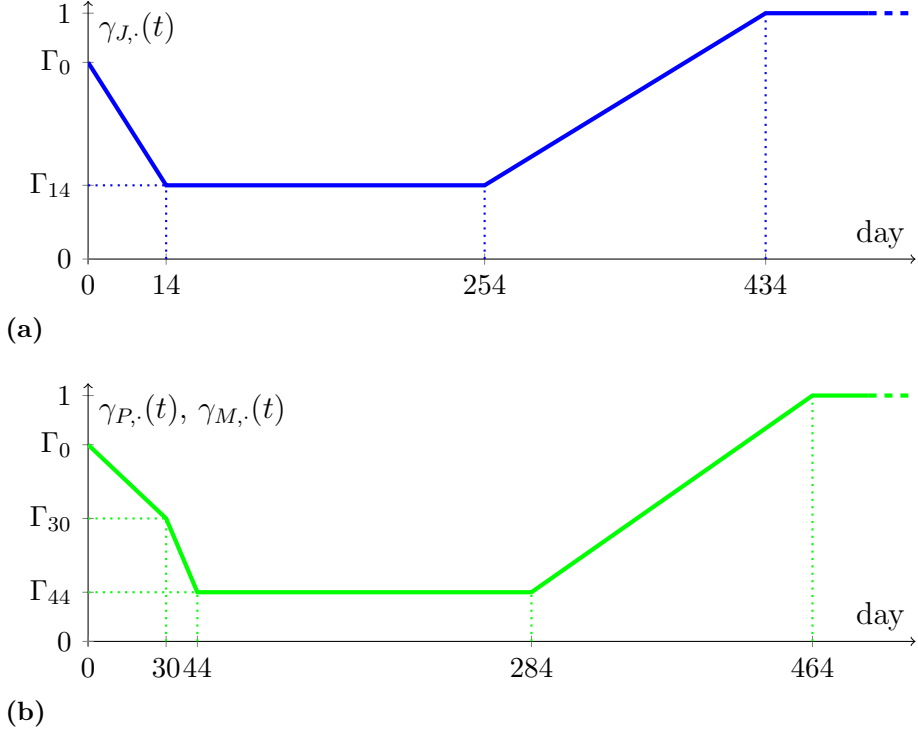

In case of two-dose vaccines, the functions have two discontinuities in their slope: one at the moment when the second shot is administered one month after the first shot (day $30$), and second two weeks (day $44$) after the second shot. Similar to the one-dose vaccine, the peak values of the functions are then kept constant for an eight-month period \cite{cdc2021booster}. Once that period is over (that is, from day $284$), the functions linearly approach $1$ over the course of six months (day $464$). Hence, here the curve is fully determined by three parameters: the value of the function immediately following the first shot ($\Gamma_0$), the value at the moment of the second shot ($\Gamma_{30}$), and the peak value ($\Gamma_{44}$), as illustrated in Fig.~\ref{fig:2dose}. These values are reported in Table~\ref{tab:vacPf} and Table~\ref{tab:vacMd} for the Pfizer and Moderna vaccine, respectively.

The characteristics of vaccination effects were based on the data officially distributed by CDC\cite{cdcJJ, cdcPfizer, cdcModerna} and the Food and Drug Administration\cite{fdaJJ, fdaPfizer, fdaModerna}. If the data for all the modeled time points was not available, we uniformly interpolated between the known values. In cases where there was only one or two datapoints, we scaled the parameter relative to one with most reported datapoints. For example, we assumed that the increase in probability of never developing symptoms, when unknown, changed proportionally to the established vaccine efficacy (that is $1-\gamma_{\cdot,1}(s)$). If no data was available, we guessed the value based on the parameter type and its relation with other vaccine benefits. In particular, we let the transmission reduction follow the drop in infection probability. We also used this relationship to extrapolate the reduction in hospitalization likelihood for the Moderna vaccine ($\gamma_{M,3}$).

We assumed that the booster shot restores peak vaccination benefits in 1 day after its administration and retains such benefits for a period that is longer than the simulation horizon (that is, six months). Hence, if an agent $i$ receives the booster shot at day $\tilde t_i$, then all the five parameters that are affected by the vaccinations detailed in the above take their peak values, that is, we set $\gamma_{J,\cdot}(t)=\Gamma_{14,\cdot}$, $\gamma_{P,\cdot}(t)=\Gamma_{44,\cdot}$, and $\gamma_{M,\cdot}(t)=\Gamma_{44,\cdot}$, for all $t>\tilde t_i$.

In this implementation, we relied on the simplifying assumption that the benefits associated with the vaccination start immediately after the administration of the shot, and then they may increase until they reach a peak value. Such a peak is attained after a fixed time interval that depends on the type of vaccine used. We found that the assumption of an immediate immunization effect of the vaccination, despite simplistic, does not qualitatively impact our  findings. Specifically, we performed a set of simulations in which initial benefits are assumed to be absent, that is, $\Gamma_0=1$ for all the entries in Tables~\ref{tab:vacJJ}--\ref{tab:vacMd}. The results of these simulations, shown in Fig.~\ref{fig:delay}, suggest that our findings are robust to (moderate) delays in the effect of vaccination.

\begin{table}
\caption{{\bf Values of the functions $\gamma$ from the one-dose Johnson\&Johnson vaccine at different times. The known  values are reported in bold with their source indicated in  brackets.}}
\centering
\begin{tabular}{|l|c|c|c|}
\hline
 & Function& $\Gamma_0$& $\Gamma_{14}$ \\
 \hline
 Infection& $\gamma_{J,1}(s)$ & {0.405} & \textbf{0.337}{~\cite{cdcJJ}}  \\
 \hline
Transmission& $\gamma_{J,2}(s)$ & 0.405 & 0.337  \\
 \hline
Hospitalization&  $\gamma_{J,3}(s)$ & \bf{0.233}$^{1}${~\cite{fdaJJ}} & \textbf{0.146}{~\cite{fdaJJ}}  \\
 \hline
Death& $\gamma_{J,4}(s)$ & 0 & \textbf{0}{~\cite{fdaJJ}} \\
 \hline
Asymptomatic&  $\gamma_{J,5}(s)$ & 1.19 & 1.326 \\
 \hline
 \end{tabular}
\label{tab:vacJJ}
 \begin{flushleft}
$^1$This value was reported after a 14 day period, but since it is lower than the peak value we use it at the moment of the vaccination.
 \end{flushleft}\end{table}

\begin{table}
\caption{{\bf Values of the functions $\gamma$ from the two-dose Pfizer vaccine at different times.} The known values are reported in bold with their source indicated in  brackets.}
\centering
\begin{tabular}{|l|c|c|c|c|}
\hline
 & Function&$\Gamma_0$& $\Gamma_{30}$ & $\Gamma_{44}$\\
 \hline
 Infection &$\gamma_{P,1}(s)$ & \textbf{0.476}{~\cite{fdaPfizer}} & \textbf{0.095}{~\cite{fdaPfizer}} & \textbf{0.05}{~\cite{fdaPfizer}} \\
 \hline
Transmission& $\gamma_{P,2}(s)$ & 0.476 & 0.095 & 0.05 \\
 \hline
Hospitalization&  $\gamma_{P,3}(s)$ & 0 & 0$^{1}$  & {0$^{1}$}\\
 \hline
Death& $\gamma_{P,4}(s)$ & 0 & {0$^{1}$} & {0$^{1}$}\\
 \hline
Asymptomatic&  $\gamma_{P,5}(s)$ & 1 & 1.524 & 1.6 \\
 \hline
 \end{tabular}
\label{tab:vacPf}
 \begin{flushleft}
{$^1$Deduced based on early CDS data in~\cite{cdcPfizer} and marginal chances of developing a severe disease reported by the FDA~\cite{fdaPfizer}.}
\end{flushleft}
\end{table}

\begin{table}
\caption{{\bf Values of the functions $\gamma$ from the two-dose Moderna vaccine at different times.} The known values are reported in bold {with their source indicated in  brackets.}}
\centering
\begin{tabular}{|l|c|c|c|c|}
\hline
 & Function& $\Gamma_0$& $\Gamma_{30}$ & $\Gamma_{44}$\\
 \hline
 Infection& $\gamma_{M,1}(s)$ & {0.25$^{1}$} & {0.154$^{1}$} & \textbf{0.059}{~\cite{fdaModerna}} \\
 \hline
Transmission& $\gamma_{M,2}(s)$ & 0.25 & 0.154 & 0.059 \\
 \hline
Hospitalization&  $\gamma_{M,3}(s)$ & 0.25 &  0.154 & \textbf{0}{~\cite{fdaModerna}} \\
 \hline
Death& $\gamma_{M,4}(s)$ & 0 & 0 & \textbf{0}{~\cite{fdaModerna}}\\
 \hline
Asymptomatic&  $\gamma_{M,5}(s)$ & 1.5 & 1.83 & 1.882 \\
 \hline
 \end{tabular}
\label{tab:vacMd}
 \begin{flushleft}
{$^1$Deduced based on early CDC data ~\cite{cdcModerna}.}
\end{flushleft}
\end{table}

\newpage

\subsection*{Out-of town non-essential locations}

As detailed in \cite{truszkowska2021designing} (Sections 4.1 and 4.3), the agents in the model can visit various non-essential locations, such as grocery stores and leisure locations. Human-to-human interactions made at these places, termed non-essential locations, contribute to the spread of the disease. In~\cite{truszkowska2021designing}, we only modeled the non-essential locations that were within the administrative limits of the town of New Rochelle. However, with the current uplifting of the lockdown measures, many residents of New Rochelle have started again visiting leisure and non-essential locations that are outside the town. To address this, in our new implementation of the model, we extended the database to include popular venues outside of town limits, as indicated by the SafeGraph data\cite{safegraph}.

In our model, the risk of infection at a location is proportional to the number of infected agents therein, as described in Section 3.2 of~\cite{truszkowska2021covid} and further detailed in Section S1 of the Supporting Information of~\cite{truszkowska2021covid}. For in-town locations, such a quantity can be exactly determined, as the model provides a one-to-one reproduction of the entire population of the town (see Section 4.1 of~\cite{truszkowska2021designing} for more details). However, this is not possible for places outside New Rochelle as it would require explicit accounting for all the people in town vicinity. Thus, we approximated the risk of infection in a out-of-town non-essential location based on the estimates on the contagion in the area in which it is located.

Following the notation introduced in Section~\ref{sec:transmission}, the infectiousness of an out-of-town non-essential location $\ell$ is defined as,
 \begin{equation}
     	\lambda_{\leisure,\ell}(t) = \beta_{\leisure}\chi_I\,,
\label{eq:lout}
 \end{equation}
where $\beta_{\leisure}$ is the transmission rate at a generic non-essential location (see Tab.~\ref{tab:parameters}), and $\chi_I$ is the COVID-19 prevalence reported for the geographic region around the town\cite{NYSCF,ConCF,USCensusQF}. 
    
\subsection{Delta variant}

To adapt the spreading to the locally dominant Delta variant, we increased transmissibility of COVID-19 by a factor of 1.6\cite{yale2021delta}. We also reduced the average latency period to 3.7 days\cite{li2021viral}. All these changes are detailed in Table~\ref{tab:parameters}.
    
\subsection{Changes in the behavior of symptomatic agents}
Infected agents with symptoms can no longer visit non-essential locations. This also holds for agents with COVID-19 like symptoms due to other diseases such as seasonal influenza. Infected agents with symptoms no longer contribute to the infection risks in public transit or carpools, which reflects their complete avoidance of other community members. 
    
\subsection{Higher education}

The age of agents who can attend higher education institutions is set to 18-24 (previously, in\cite{truszkowska2021covid,truszkowska2021designing}, it was 18-21).

\subsection{Public transit routes}

Transit routes and their destinations for agents traveling to work in and outside of New Rochelle are reported in Table~\ref{tab:route}.

\subsection{Details about the setup of vaccinated individuals}

As of September 7\textsuperscript{th} 2021, 51,342 individuals received  the first shot of the vaccine\cite{NYSTVac}. We initialized these individuals by setting the date of administration of their first shot to a random time between January 1\textsuperscript{th} 2021 and September 6\textsuperscript{th} 2021. Hence, we implicitly assumed that the rate at which individuals received the first shot was constant  --this is illustrated in Fig.~\ref{fig:day}. The types of  vaccines administered to these individuals mirrored those that were distributed in the area and included the two double-dose vaccines (Moderna and Pfizer) and one single-dose vaccine (Johnson\&Johnson), see Tab.~\ref{tab:vacProps}. 

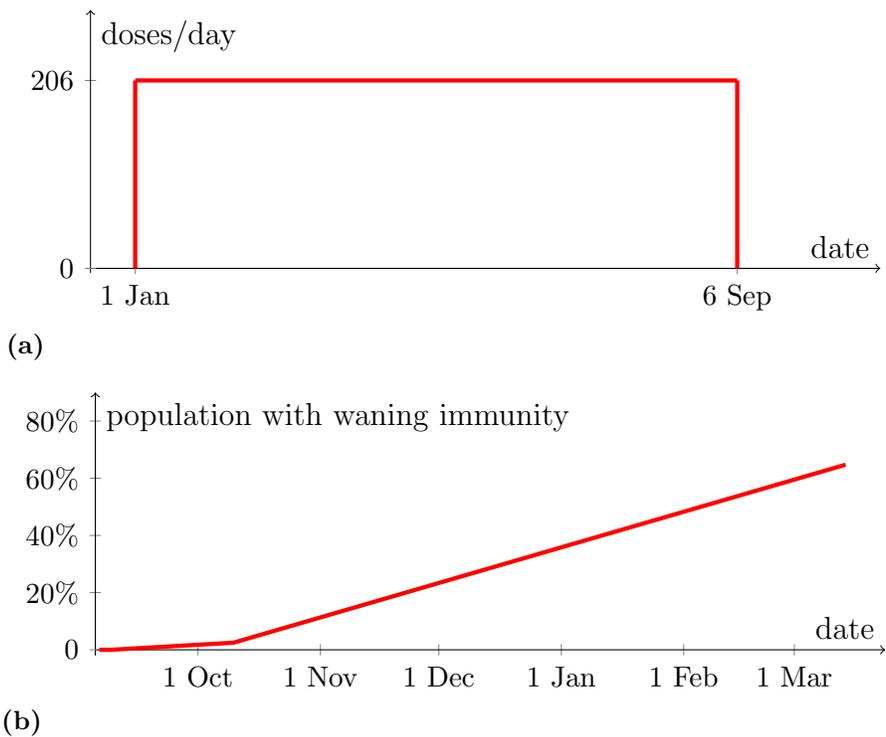
\begin{figure}
    \centering
  \subfloat[]{ \input{day}    \label{fig:day}}  \\
  \subfloat[]{ \input{day_wane}    \label{fig:day_wane}}
    \caption{{\bf (a)} Temporal profile of first shots used for the initialization of the simulations, and {\bf (b)} cumulative fraction of initially vaccinated population that starts experiencing waning immunity during our six-month simulations. }

\end{figure}

According to the assumed vaccination profile prior to the start of our study, none of the vaccinated individuals experience waning immunity as of September 6\textsuperscript{th} 2021; whereas 65\% of them should experience waning immunity during the course of the six-month simulation  -- the subject of our study. The cumulative fraction of the initially vaccinated population that starts experiencing waning immunity is shown in Fig.~\ref{fig:day_wane}.

Note that precise data on vaccination timing in the town of New Rochelle and on the different types of vaccine used are not available, as they are only reported at the county level and in an aggregate form~\cite{vacTrack}. For this reason, we preferred to perform our numerical simulations by assuming uniform distributions to emulate their administration. In Fig.~\ref{fig:distribution}, we illustrate the robustness of our qualitative findings with respect to the temporal profile of the first shot. Specifically, we scale the available county-level aggregate data~\cite{vacStats} to estimate the temporal distribution of first shots in the town of New Rochelle, and we use the  obtained distribution to initialize our simulations. Results of our simulations do not show major differences with respect to those obtained using a uniform distribution, illustrated in Fig.~3 in the main article.

\newpage

\section{Model parameters}

\begin{longtable}[ht!]{|m{6cm}|m{4cm}|m{4cm}|}
\caption{{\bf Other parameters of the ABM.}  $^1$ Scaled down to town size and time-step. $^2$  Scaled down to town size, time-step, and doubled following calibrated percentage of asymptomatic adults in~Ref.~\cite{truszkowska2021covid}, used as a proxy for underdetection. $^3$ Scaled down to town size, time-step, and doubled following calibrated percentage of asymptomatic adults in~Ref.~\cite{truszkowska2021covid}, used as a proxy for underdetection; this is the total number of cases recovering from COVID-19 during an average recovery period used in~Ref.~\cite{truszkowska2021covid} and scaled with the new latency duration. $^4$ Based on decreasing numbers of new vaccinations in New York State~\cite{OWD_us_vac}.\label{S1_Table}}\\
\hline
& \textbf{Value}& \textbf{Reference} \\
 \hline
 Increase of all COVID-19 transmission rates due to Delta variant & 1.6 & \cite{yale2021delta}\\
 \hline
 Fraction of the population that is estimated to be infected in the area at a time-step & 0.0003 & \cite{NYSCF,ConCF,USCensusQF}\\
 \hline
Transmission rate in a out-of-town workplace  & 0.000318 & \cite{NYSCF,ConCF,USCensusQF}\\
 \hline
Transmission rate in a out-of-town leisure location  & 0.00010944 & \cite{NYSCF,ConCF,USCensusQF}\\
\hline
 Current capacity of public transit compared to its maximum capacity & 0.66 & \cite{GoogleMRSept} for public transit\\
 \hline
 Fraction of susceptible agents with COVID-19-like symptoms & 1e-6 &  \cite{CDCFlu} \\ 
 \hline
  Latency period & log-normal distribution with 1.225 mean and 0.418 standard deviation, $\mathrm{days}$ & \cite{lauer2020incubation,li2021viral} \\
   \hline
 Fraction of the nominal transmission rate at workplaces, public transit, carpools, and leisure locations associated with current reopening efforts & 0.62 &  \cite{GoogleMRSept} for workplaces\\
  \hline
 Fraction of agents going to leisure locations at each time-step& 0.5 & Assumption\\
  \hline
 Fraction of fully vaccinated agents going to leisure locations at each time-step& 0.75 & Assumption\\
  \hline
Total initial population &79,205 & \cite{USCensus}\\
 \hline
 Initial number of vaccinated agents & 51,342 & \cite{NYSTVac}$^1$\\
  \hline
 Maximum number of agents that can be vaccinated & 64,364 & Assumption$^4$\\
 \hline
 Time before recovery and vaccination eligibility & 21 days & Assumption\\
  \hline
    Duration of natural immunity after recovery & 180 days & \cite{Gudbjartsson2020immunity,Dan2021immunity}\\
  \hline
    Compliance to home isolation after potential exposure from a house guest (contact tracing) & 0.109 & \cite{smith2021adherence}\\
  \hline
    Maximum number of coworkers, students, or retirement home residents traced & 10 & Assumption\\
  \hline
    Duration of home-isolation & 10 days & \cite{cdc2021ctracing,NYSCT,NRCT}\\
  \hline
Duration of the awareness after home-isolation & 4 days & \cite{cdc2021ctracing,NYSCT,NRCT}\\
 \hline
 Number of initially infected agents in the town& 4 & \cite{NYSCF}\\ 
 \hline
 Number of agents that are initially active COVID-19 cases & 66 & \cite{OWD_us}$^3$\\
 \hline
 \label{tab:parameters}
\end{longtable}

\begin{table}
\centering
\caption{\bf Parameters related to vaccinations and booster campaign.\label{tab:vacProps}}\vspace{.5cm}

\begin{tabular}{|m{6cm}|l|l|}
\hline
 & \textbf{Value}& \textbf{Reference} \\
 \hline
 Fraction of people taking Johnson\&Johnson vaccine & 20\% & \cite{vacLocal}\\
 \hline
 Fraction of people taking Pfeizer vaccine & 45\% & \cite{vacLocal}\\
 \hline
 Fraction of people taking Moderna vaccine & 35\% & \cite{vacLocal}\\
 \hline
 Minimum vaccination age & 12 years old & \cite{vacFAQ}\\
 \hline
 Start of the vaccination campaign & January 1\textsuperscript{st} 2021 & Assumption\\
 \hline
 Time for the booster to restore the peak benefits & 1 day after the shot& Assumption\\
 \hline
 Duration of booster effects & 240 days after the shot& Assumption\\
 \hline
 Complete end of booster effects & 420 days after the shot& Assumption\\
 \hline
 \end{tabular}
\end{table}

\begin{table}
\centering
\caption{\bf Transit routes and their destinations for agents traveling to work in and outside of New Rochelle.}\vspace{.5cm}

\begin{tabular}{|m{7cm}|m{7cm}|}
\hline
 \textbf{Route name} & \textbf{Destination (Postal code or NR if in New Rochelle)}\\
 \hline
 Bus 0007 & NR, 10710, 10704, 10552, 10532\\
 \hline
 Bus 0040 & 10601\\
 \hline
 New Haven MTA - South direction & 10019, 10017, 10022, 10550, 10001, 10036, 10467, 10701, 10018, 10016, 10013, 10003, 11201, 10010, 10011, 11101, 10065, 10004, 10027, 10458, 10474, 10005, 10029, 10457, 10023, 10463, 0471, 10468, 1030, 11375, 10007, 10038, 10462, 10021, 11747, 10014, 10032, 10705, 10451, 10012, 10460, 10456, 10035, 10020, 10025, 10167, 10455, 10006, 10024, 10033, 10028, 10452 \\
 \hline
 New Haven MTA - North-East direction & 10543, 10573, 10580, 6830, 6831, 6901 \\
  \hline
 Bus 0045 & NR, 10709 \\
 \hline
 Bus 0060 & NR, 10604, 10466, 10595 \\
 \hline
  New Haven MTA - North direction & 10605, 10591, 10577, 10528, 10603, 10530, 10570, 10562, 10549, 12601\\
 \hline
 Bus 0042 & NR, 10553, 10470\\
 \hline
 Bus 0045 Southbound & 10461, 10465, 10707 \\
 \hline
 Bus 0006 & 10595 \\
 \hline
 Bus 0030 & 10708, 10703\\
 \hline
 Amtrak- Northeast Regional & 06902\\
 \hline
 Bus 0062 & 10523, 10606, 10504, 10954\\
 \hline
 Bus 0061 & 10475, 10469 \\
 \hline
 Bus 0066 & 10522 \\
 \hline
 Amtrak- Northeast Regional - Southbound & 11530, 11581\\
 \hline
 \end{tabular}\label{tab:route}
\end{table}

\pagebreak
\newpage
	\pagestyle{empty}

\section{Additional simulations}

In the next pages, we report here some additional simulations to test robustness of our findings.

\begin{figure}
 \centering
   \includegraphics[width=\textwidth]{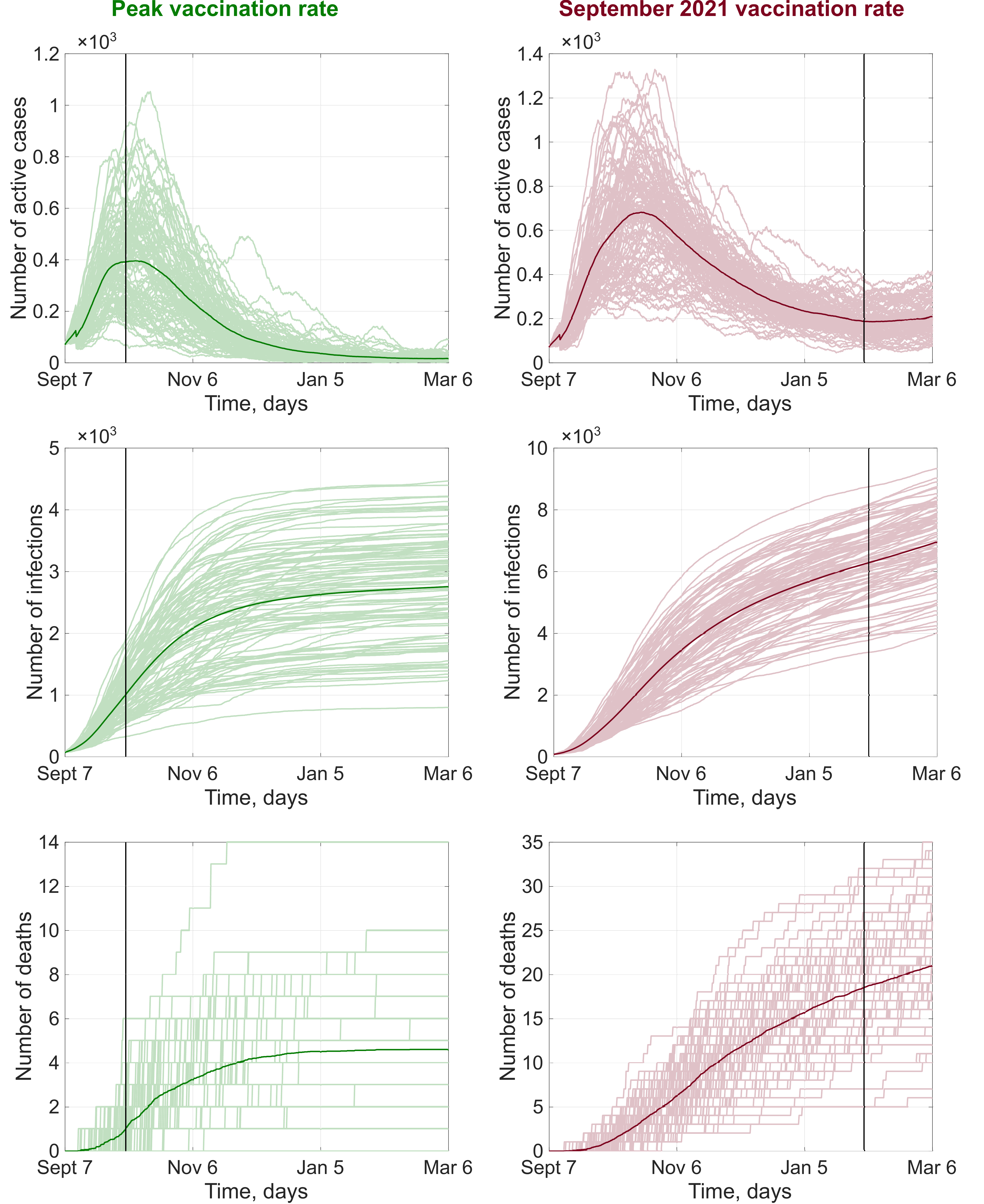}
   \caption{{\bf COVID-19 spreading over six months from September 7\textsuperscript{th} 2021, amid two different vaccination campaigns, assuming that vaccine immunity starts to wane after eight months and natural immunity lasts four months.} Active cases, total number of infections, and total deaths for the next six months at either peak vaccination rate of 0.58\% population/day  (green) or present vaccination rate of 0.11\% population/day (red). For each scenario, $100$ independent realizations are shown and their average is highlighted. The vertical lines denote the date at which the entire non-hesitant eligible population is expected to be vaccinated with at least one shot. The simulations in this figure are consistent with those in Fig.~3, where natural immunity lasts six months, thus corroborating the robustness of our findings. }
   \label{fig:4month_nat}
\end{figure}

\begin{figure}
 \centering
   \includegraphics[width=\textwidth]{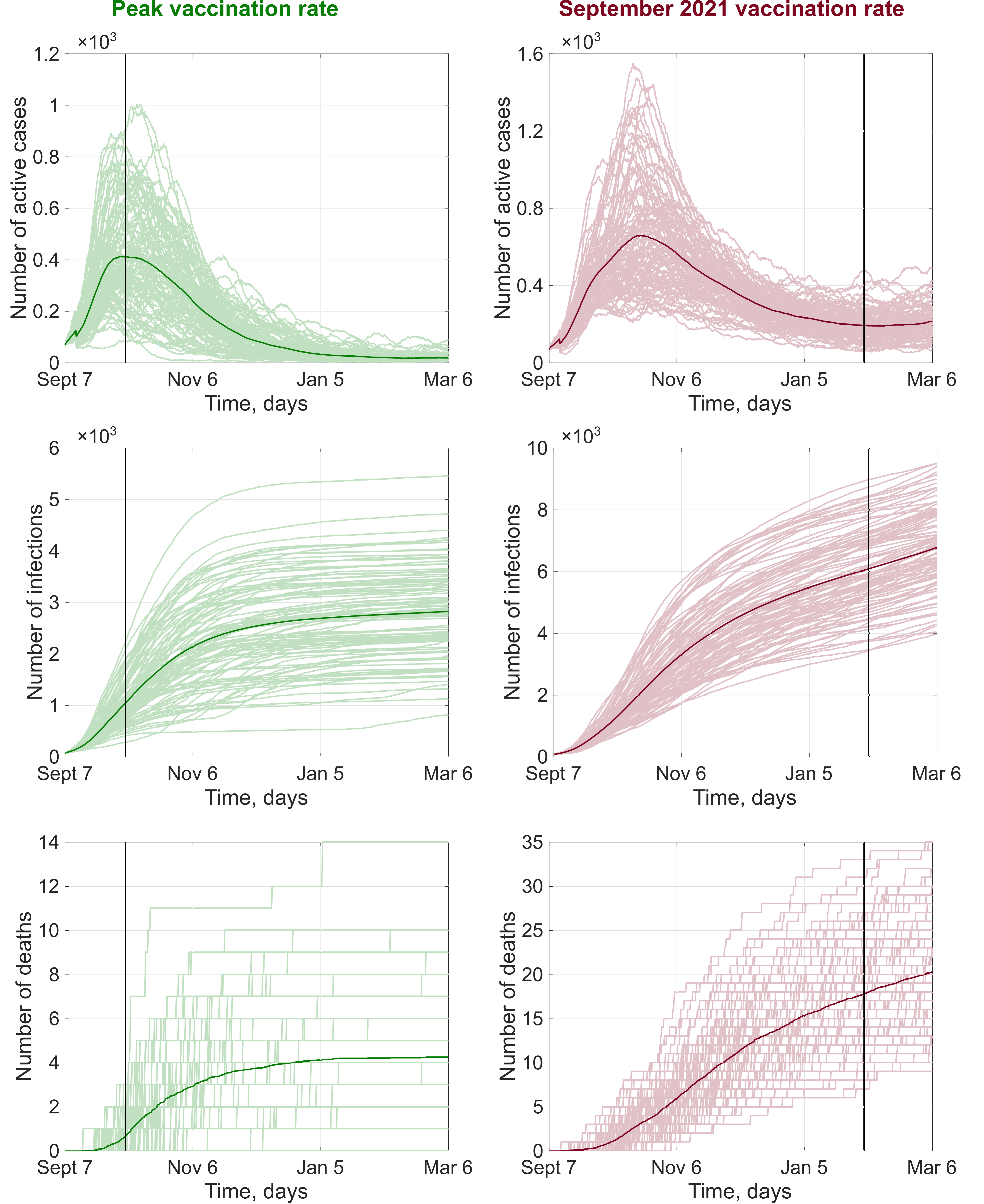}
   \caption{{\bf COVID-19 spreading over six months from September 7\textsuperscript{th} 2021, amid two different vaccination campaigns, assuming that vaccine immunity starts to wane after eight months and natural immunity lasts eight months.} Active cases, total number of infections, and total deaths for the next six months at either peak vaccination rate of 0.58\% population/day  (green) or present vaccination rate of 0.11\% population/day (red). For each scenario, $100$ independent realizations are shown and their average is highlighted. The vertical lines denote the date at which the entire non-hesitant eligible population is expected to be vaccinated with at least one shot. The simulations in this figure are consistent with those in Fig.~3, where natural immunity lasts six months, thus corroborating the robustness of our findings. }
   \label{fig:8month_nat}
\end{figure}

\begin{figure}
 \centering
   \includegraphics[width=.9\textwidth]{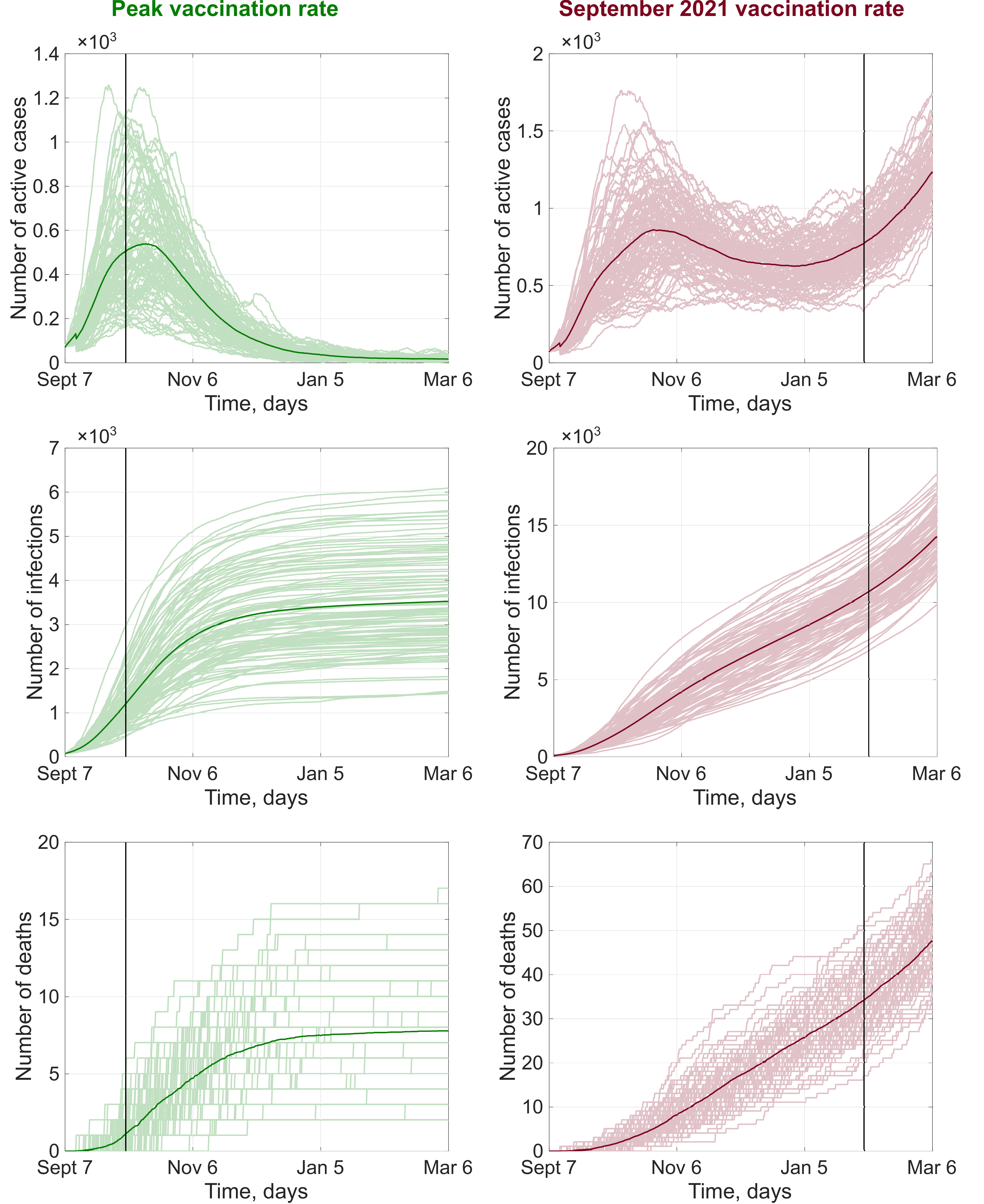}
   \caption{{\bf COVID-19 spreading over six months from September 7\textsuperscript{th} 2021, amid two different vaccination campaigns, assuming that vaccine immunity starts to wane after six months and natural immunity lasts six months.} Active cases, total number of infections, and total deaths for the following six months at either peak vaccination rate of 0.58\% population/day (green) or the present vaccination rate of 0.11\% population/day (red). For each scenario, $100$ independent realizations are shown and their average is highlighted. The vertical lines denote the date at which the entire non-hesitant eligible population is expected to be vaccinated with at least one shot. Anticipating the time at which immunity starts to wane leads to worse outcomes when the vaccination rate is low (red) compared to the original simulations in Fig.~3 in the main article; while it has no effect for high vaccination rates. These simulations  confirm our findings on the need of high vaccination rates to eradicate COVID-19. }
   \label{fig:6month_s}
\end{figure}

\begin{figure}
 \centering
   \includegraphics[width=.9\textwidth]{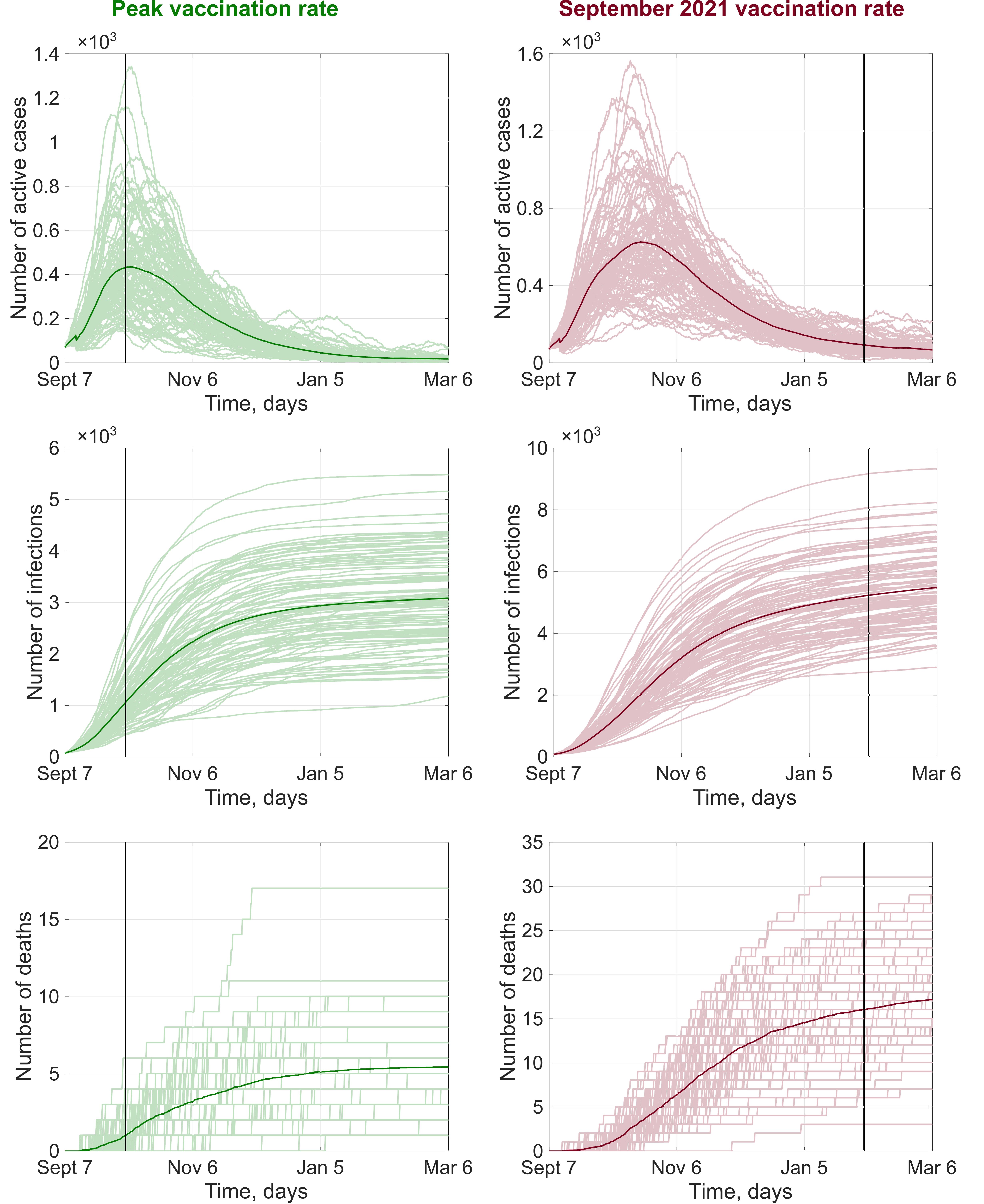}
   \caption{{\bf COVID-19 spreading over six months from September 7\textsuperscript{th} 2021, amid two different vaccination campaigns, assuming that vaccine immunity starts to wane after 10 months and natural immunity lasts six months.} Active cases, total number of infections, and total deaths for the next six months at either peak vaccination rate of 0.58\% population/day (green) or present vaccination rate of 0.11\% (red) population/day. For each scenario, $100$ independent realizations are shown and their average is highlighted. The vertical lines denote the date at which the entire non-hesitant eligible population is expected to be vaccinated with at least one shot. Predictably, delaying the time at which immunity starts to wane leads to better outcomes even when the vaccination rate is low (red). However, this improvement seems to be not sufficient to completely eradicate the disease, differently from what observed for high vaccination rate (green), whose outcomes remain almost unchanged. These observations confirm our findings on the need of high vaccination rates to eradicate COVID-19. }
   \label{fig:10month}
\end{figure}

\begin{figure}
 \centering
   \includegraphics[width=\textwidth]{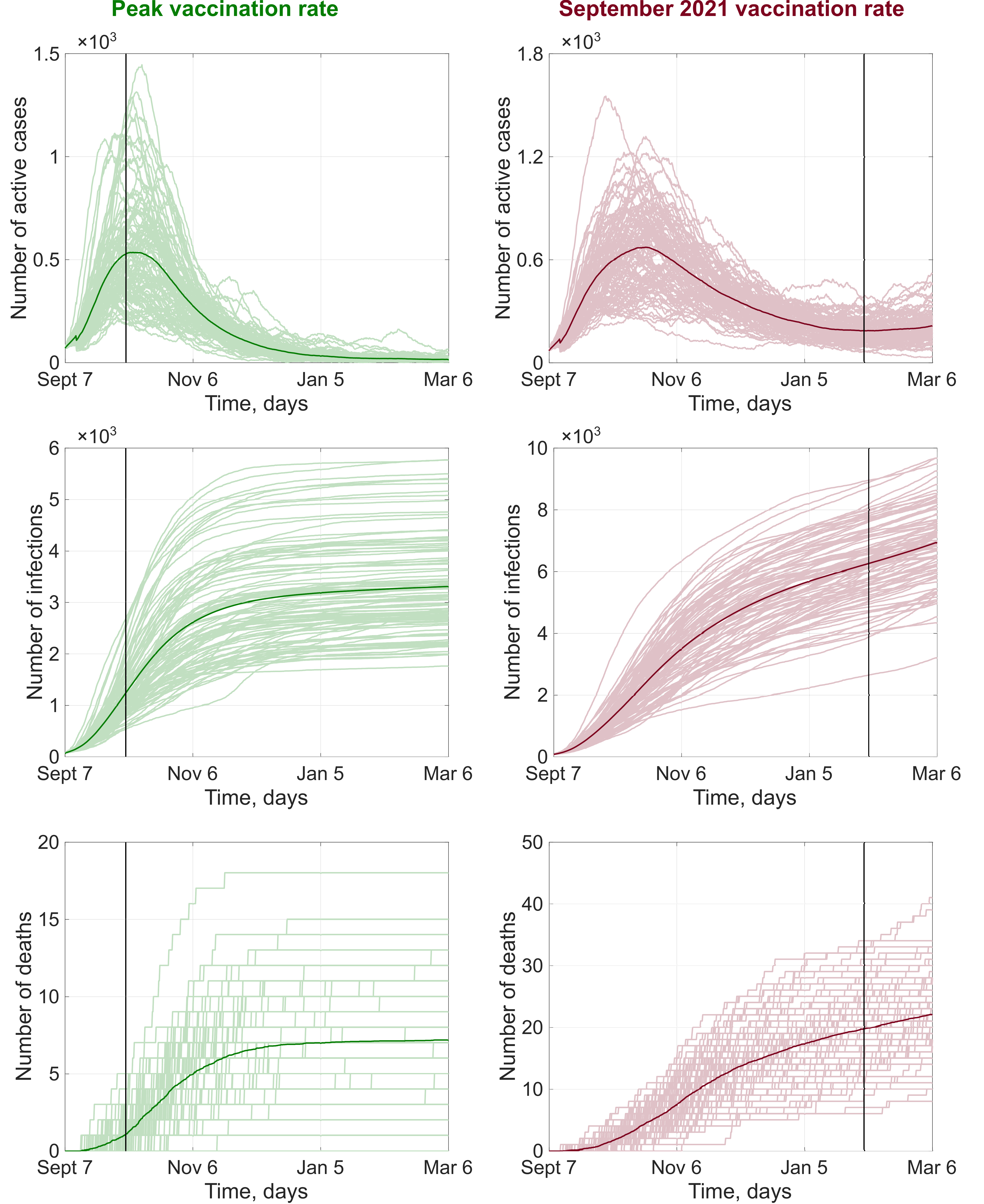}
   \caption{{\bf COVID-19 spreading over six months from September 7\textsuperscript{th} 2021, with delayed vaccination benefits, that is, with $\Gamma_0=1$.} Active cases, total number of infections, and total deaths for the next six months at either peak vaccination rate of 0.58\% population/day  (green) or present vaccination rate of 0.11\% population/day (red). For each scenario, $100$ independent realizations are shown and their average is highlighted. The vertical lines denote the date at which the entire non-hesitant eligible population is expected to be vaccinated with at least one shot. The results preserve the trends observed in the simulations in the main paper and do not show major quantitative differences, indicating robustness of our numerical findings.}
   \label{fig:delay}
\end{figure}

\begin{figure}
 \centering
   \includegraphics[width=\textwidth]{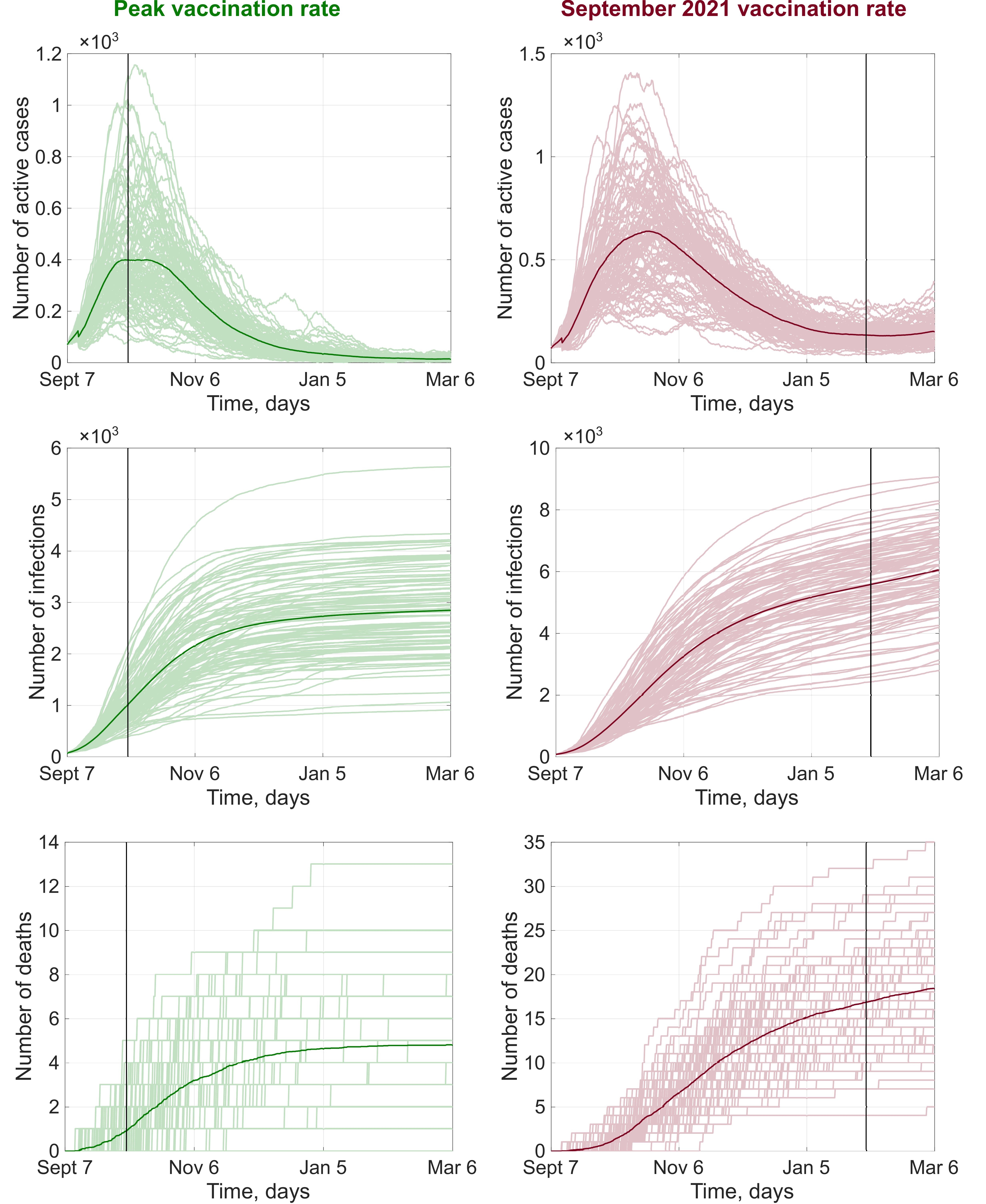}
   \caption{{\bf COVID-19 spreading over six months from September 7\textsuperscript{th} 2021, with temporal profile of first shots estimated from available aggregate data at the county level.} Active cases, total number of infections, and total deaths for the next six months at either peak vaccination rate of 0.58\% population/day  (green) or present vaccination rate of 0.11\% population/day (red). For each scenario, $100$ independent realizations are shown and their average is highlighted. The vertical lines denote the date at which the entire non-hesitant eligible population is expected to be vaccinated with at least one shot. The results do not show significant differences with ones in Fig.~3, indicating robustness of our numerical findings obtained under assumption of uniform distribution of vaccinations.}
   \label{fig:distribution}
\end{figure}

\begin{figure}
 \centering
   \includegraphics[width=.9\textwidth]{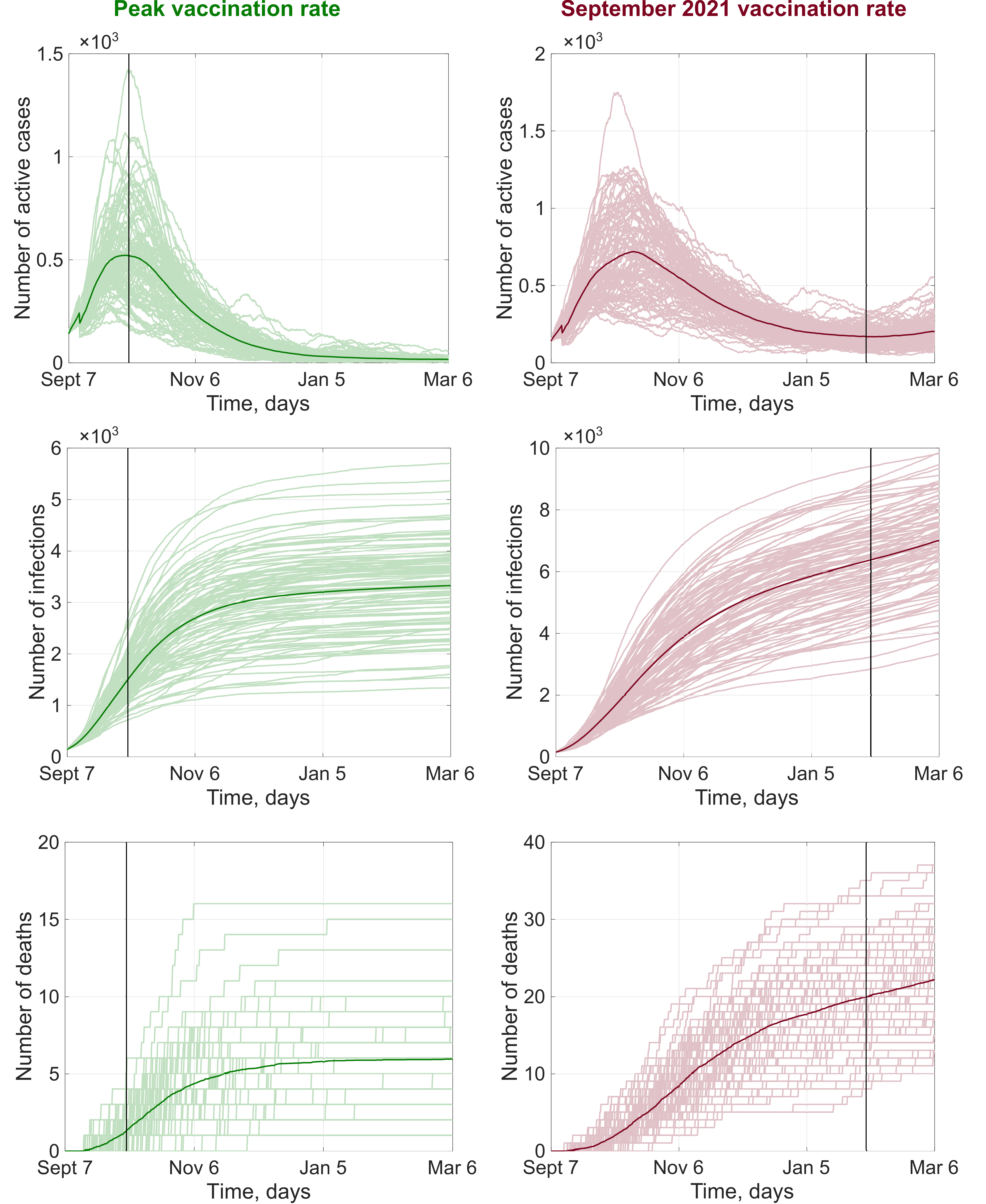}
   \caption{{\bf COVID-19 spreading over six months from September 7\textsuperscript{th} 2021, assuming that only $50\%$ of the initially infections are reported.} Active cases, total number of infections, and total deaths for the next six months at either peak vaccination rate of 0.58\% population/day  (green) or present vaccination rate of 0.11\% population/day (red). For each scenario, $100$ independent realizations are shown and their average is highlighted. The vertical lines denote the date at which the entire non-hesitant eligible population is expected to be vaccinated with at least one shot. As expected, the different initialization of the simulations impacts the initial phase of the spreading process, yielding a sliglty higher COVID-19 toll and anticipating the peak with respect to the simulations in Fig.~3. However, such an effect is observed uniformly across all the simulations and its impact on the overall outcome of the simulations and on the comparison between the different scenarios is negligible. }
   \label{fig:doubleI}
\end{figure}

\begin{figure}
 \centering
   \includegraphics[width=.9\textwidth]{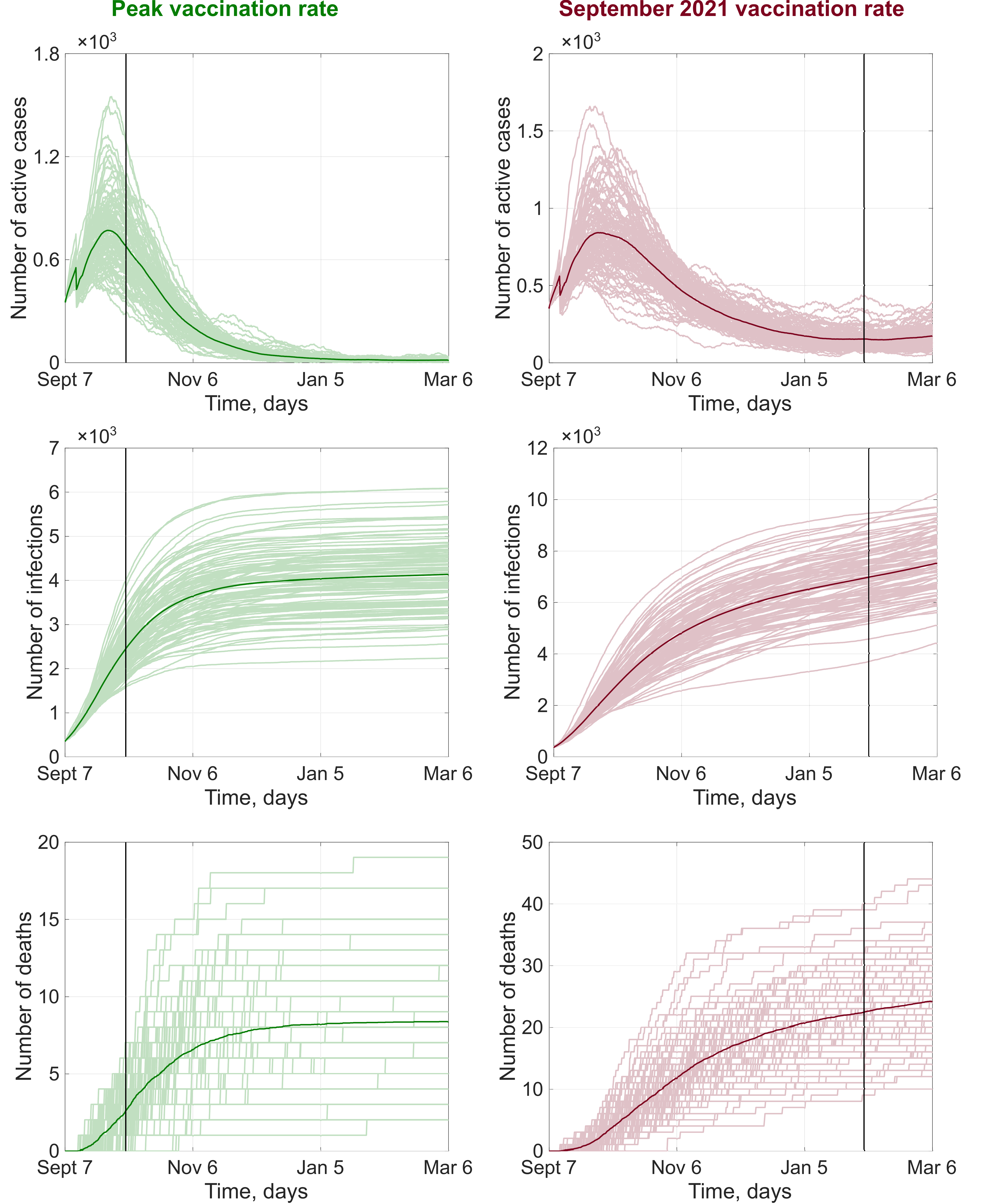}
   \caption{{\bf COVID-19 spreading over six months from September 7\textsuperscript{th} 2021, assuming that only $20\%$ of the initially infections are reported.} Active cases, total number of infections, and total deaths for the next six months at either peak vaccination rate of 0.58\% population/day  (green) or present vaccination rate of 0.11\% population/day (red). For each scenario, $100$ independent realizations are shown and their average is highlighted. The vertical lines denote the date at which the entire non-hesitant eligible population is expected to be vaccinated with at least one shot. Predictably, the different initialization affects the initial phase of the spreading process, increasing the overall COVID-19 toll and shifting the peak with respect to the simulations in Fig.~3. However, similar to what has been observed in Fig.~\ref{fig:doubleI}, such an effect is uniform across all the simulations and thus the impact of the initial condition on our qualitative numerical findings and on the comparison between the different scenarios is negligible. }
   \label{fig:fiveI}
\end{figure}

\begin{figure}
 \centering
   \includegraphics[width=.95\textwidth]{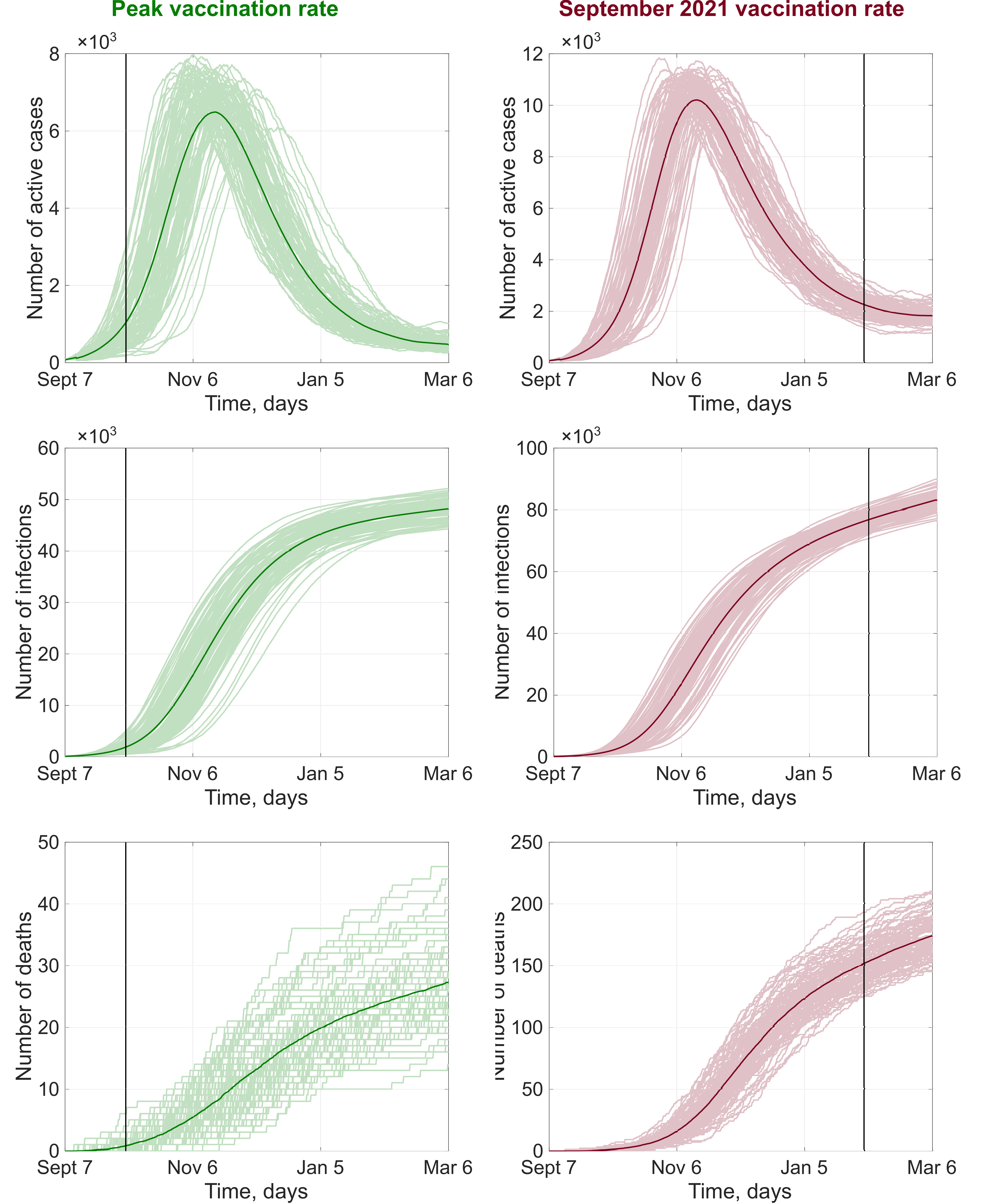}
   \caption{{\bf COVID-19 spreading over six months from September 7\textsuperscript{th} 2021, without any form of contact tracing.} Active cases, total number of infections, and total deaths for the next six months at either peak vaccination rate of 0.58\% population/day  (green) or present vaccination rate of 0.11\% population/day (red). For each scenario, $100$ independent realizations are shown and their average is highlighted. The vertical lines denote the date at which the entire non-hesitant eligible population is expected to be vaccinated with at least one shot. The outcomes are dramatically worse than the one in the presence of contact tracing shown in Fig.~3. Even in the scenario of peak vaccination rate, the absence of any form of contact tracing would lead to a more than ten-fold increase in the cumulative number of infections and more than five-fold increase in the death toll.}
   \label{fig:nct}
\end{figure}

\end{document}

%% file: 1dose.tex
\begin{tikzpicture}
\begin{axis}[%
 axis lines=middle,
 x   axis line style={->},
  y   axis line style={->},
width=11 cm,
height=3.5 cm,
scale only axis,
xmin=0,
xmax=530,
     ytick ={0,.3,.8,1},
     yticklabels={$0$, $\Gamma_{14}$, $\Gamma_0$,1},
     xtick ={0,50, 254, 434},
     xticklabels={$0$, $14$, $254$, $434$},
      extra y ticks ={0},
    extra y tick labels={$0$},
     extra x ticks ={0},
     extra x tick labels={$0$},
xlabel={day},
ylabel={$\gamma_{J,\cdot}(t)$},
ymin=-0.02,
      ytick distance={0.5},
ymax=1.05,
legend style={at={(.5,1)},legend cell align=left, align=left, draw=none,fill=none, font=\small},
axis background/.style={fill=white},
]
\addplot[ultra thick, blue]
  table[row sep=crcr]{%
0 0.8\\
50 0.3\\
254 0.3\\
434 1\\
500 1\\
};

\addplot[ultra thick, dashed,blue]
  table[row sep=crcr]{%
500 1\\
530 1\\
};

\addplot[dotted, thick, blue]
  table[row sep=crcr]{%
0 0.3\\
50 0.3\\
};
\addplot[dotted, thick, blue]
  table[row sep=crcr]{%
50  0\\
50 0.3\\
};

\addplot[dotted, thick, blue]
  table[row sep=crcr]{%
254  0\\
254 0.3\\
};

\addplot[dotted, thick, blue]
  table[row sep=crcr]{%
434  0\\
434 1\\
};

\end{axis}
\end{tikzpicture}

%% file: 2dose.tex
\begin{tikzpicture}
\begin{axis}[%
 axis lines=middle,
 x   axis line style={->},
  y   axis line style={->},
width=11 cm,
height=3.5 cm,
scale only axis,
xmin=0,
xmax=530,
     ytick ={0,.2,.5, .8,1},
     yticklabels={$0$, $\Gamma_{44}$, $\Gamma_{30}$, $\Gamma_0$,1},
     xtick ={0,50, 70, 284, 464},
     xticklabels={$0$, $30$, $44$, $284$, $464$},
      extra y ticks ={0},
    extra y tick labels={$0$},
     extra x ticks ={0},
     extra x tick labels={$0$},
xlabel={day},
ylabel={$\gamma_{P,\cdot}(t)$, $\gamma_{M,\cdot}(t)$},
ymin=-0.02,
      ytick distance={0.5},
ymax=1.05,
legend style={at={(.5,1)},legend cell align=left, align=left, draw=none,fill=none, font=\small},
axis background/.style={fill=white},
]
\addplot[ultra thick, green]
  table[row sep=crcr]{%
0 0.8\\
50 0.5\\
70 0.2\\
284 0.2\\
464 1\\
500 1\\
};

\addplot[ultra thick, dashed,green]
  table[row sep=crcr]{%
500 1\\
530 1\\
};

\addplot[dotted, thick, green]
  table[row sep=crcr]{%
0 0.5\\
50 0.5\\
};

\addplot[dotted, thick, green]
  table[row sep=crcr]{%
0 0.2\\
70 0.2\\
};

\addplot[dotted, thick, green]
  table[row sep=crcr]{%
50 0\\
50 0.5\\
};

\addplot[dotted, thick, green]
  table[row sep=crcr]{%
70 0\\
70 0.2\\
};

\addplot[dotted, thick, green]
  table[row sep=crcr]{%
284 0\\
284 0.2\\
};

\addplot[dotted, thick, green]
  table[row sep=crcr]{%
464 0\\
464 1\\
};

\end{axis}
\end{tikzpicture}

%% file: day.tex
\begin{tikzpicture}
\begin{axis}[%
 axis lines=middle,
 x   axis line style={->},
  y   axis line style={->},
width=10.5 cm,
height=3.5 cm,
scale only axis,
xmin=0,
xmax=530,
     ytick ={0,.8},
     yticklabels={$0$, $206$},
     xtick ={30, 434},
     xticklabels={1 Jan, 6 Sep},
      extra y ticks ={0},
    extra y tick labels={$0$},
xlabel={date},
ylabel={doses/day},
ymin=-0.02,
      ytick distance={0.5},
ymax=1.1,
legend style={at={(.5,1)},legend cell align=left, align=left, draw=none,fill=none, font=\small},
axis background/.style={fill=white},
]
\addplot[ultra thick, red]
  table[row sep=crcr]{%
30 0.8\\
434 0.8\\
};

\addplot[ultra thick, red]
  table[row sep=crcr]{%
434 0\\
434 0.8\\
};

\addplot[ultra thick, red]
  table[row sep=crcr]{%
30 0\\
30 0.8\\
};

\end{axis}
\end{tikzpicture}

%% file: day_wane.tex
\begin{tikzpicture}
\begin{axis}[%
 axis lines=middle,
 x   axis line style={->},
  y   axis line style={->},
width=10.5 cm,
height=3.5 cm,
scale only axis,
xmin=0,
xmax=200,
     ytick ={0,.2,.4,.6,.8},
     yticklabels={$0\%$,$20\%$,$40\%$,$60\%$,$80\%$},
     xtick ={26, 57,87,118,149,177},
     xticklabels={1 Oct, 1 Nov, 1 Dec, 1 Jan, 1 Feb, 1 Mar},
      extra y ticks ={0},
    extra y tick labels={$0$},
xlabel={date},
ylabel={population with waning immunity},
ymin=-0.02,
      ytick distance={0.5},
ymax=.9,
legend style={at={(.5,1)},legend cell align=left, align=left, draw=none,fill=none, font=\small},
axis background/.style={fill=white},
]
\addplot[ultra thick, red]
  table[row sep=crcr]{%
1	0\\
2	0\\
3	0\\
4	0\\
5	0.000803212851405623\\
6	0.00160642570281125\\
7	0.00240963855421687\\
8	0.00321285140562249\\
9	0.00401606425702811\\
10	0.00481927710843373\\
11	0.00562248995983936\\
12	0.00642570281124498\\
13	0.0072289156626506\\
14	0.00803212851405622\\
15	0.00883534136546185\\
16	0.00963855421686747\\
17	0.0104417670682731\\
18	0.0112449799196787\\
19	0.0120481927710843\\
20	0.01285140562249\\
21	0.0136546184738956\\
22	0.0144578313253012\\
23	0.0152610441767068\\
24	0.0160642570281125\\
25	0.0168674698795181\\
26	0.0176706827309237\\
27	0.0184738955823293\\
28	0.0192771084337349\\
29	0.0200803212851406\\
30	0.0208835341365462\\
31	0.0216867469879518\\
32	0.0224899598393574\\
33	0.0232931726907631\\
34	0.0240963855421687\\
35	0.0248995983935743\\
36	0.0289156626506024\\
37	0.0329317269076305\\
38	0.0369477911646587\\
39	0.0409638554216868\\
40	0.0449799196787149\\
41	0.048995983935743\\
42	0.0530120481927711\\
43	0.0570281124497992\\
44	0.0610441767068273\\
45	0.0650602409638554\\
46	0.0690763052208836\\
47	0.0730923694779117\\
48	0.0771084337349398\\
49	0.0811244979919679\\
50	0.085140562248996\\
51	0.0891566265060241\\
52	0.0931726907630522\\
53	0.0971887550200804\\
54	0.101204819277108\\
55	0.105220883534137\\
56	0.109236947791165\\
57	0.113253012048193\\
58	0.117269076305221\\
59	0.121285140562249\\
60	0.125301204819277\\
61	0.129317269076305\\
62	0.133333333333333\\
63	0.137349397590361\\
64	0.14136546184739\\
65	0.145381526104418\\
66	0.149397590361446\\
67	0.153413654618474\\
68	0.157429718875502\\
69	0.16144578313253\\
70	0.165461847389558\\
71	0.169477911646586\\
72	0.173493975903615\\
73	0.177510040160643\\
74	0.181526104417671\\
75	0.185542168674699\\
76	0.189558232931727\\
77	0.193574297188755\\
78	0.197590361445783\\
79	0.201606425702811\\
80	0.205622489959839\\
81	0.209638554216868\\
82	0.213654618473896\\
83	0.217670682730924\\
84	0.221686746987952\\
85	0.22570281124498\\
86	0.229718875502008\\
87	0.233734939759036\\
88	0.237751004016064\\
89	0.241767068273092\\
90	0.24578313253012\\
91	0.249799196787149\\
92	0.253815261044177\\
93	0.257831325301205\\
94	0.261847389558233\\
95	0.265863453815261\\
96	0.269879518072289\\
97	0.273895582329317\\
98	0.277911646586345\\
99	0.281927710843373\\
100	0.285943775100401\\
101	0.289959839357429\\
102	0.293975903614458\\
103	0.297991967871486\\
104	0.302008032128514\\
105	0.306024096385542\\
106	0.31004016064257\\
107	0.314056224899598\\
108	0.318072289156626\\
109	0.322088353413654\\
110	0.326104417670682\\
111	0.33012048192771\\
112	0.334136546184738\\
113	0.338152610441767\\
114	0.342168674698795\\
115	0.346184738955823\\
116	0.350200803212851\\
117	0.354216867469879\\
118	0.358232931726907\\
119	0.362248995983935\\
120	0.366265060240963\\
121	0.370281124497991\\
122	0.374297188755019\\
123	0.378313253012048\\
124	0.382329317269076\\
125	0.386345381526104\\
126	0.390361445783132\\
127	0.39437751004016\\
128	0.398393574297188\\
129	0.402409638554216\\
130	0.406425702811244\\
131	0.410441767068272\\
132	0.4144578313253\\
133	0.418473895582329\\
134	0.422489959839357\\
135	0.426506024096385\\
136	0.430522088353413\\
137	0.434538152610441\\
138	0.438554216867469\\
139	0.442570281124497\\
140	0.446586345381525\\
141	0.450602409638553\\
142	0.454618473895581\\
143	0.458634538152609\\
144	0.462650602409638\\
145	0.466666666666666\\
146	0.470682730923694\\
147	0.474698795180722\\
148	0.47871485943775\\
149	0.482730923694778\\
150	0.486746987951806\\
151	0.490763052208834\\
152	0.494779116465862\\
153	0.49879518072289\\
154	0.502811244979918\\
155	0.506827309236947\\
156	0.510843373493975\\
157	0.514859437751003\\
158	0.518875502008031\\
159	0.522891566265059\\
160	0.526907630522087\\
161	0.530923694779115\\
162	0.534939759036143\\
163	0.538955823293171\\
164	0.542971887550199\\
165	0.546987951807228\\
166	0.551004016064256\\
167	0.555020080321284\\
168	0.559036144578312\\
169	0.56305220883534\\
170	0.567068273092368\\
171	0.571084337349396\\
172	0.575100401606424\\
173	0.579116465863452\\
174	0.58313253012048\\
175	0.587148594377509\\
176	0.591164658634537\\
177	0.595180722891565\\
178	0.599196787148593\\
179	0.603212851405621\\
180	0.607228915662649\\
181	0.611244979919677\\
182	0.615261044176705\\
183	0.619277108433733\\
184	0.623293172690761\\
185	0.62730923694779\\
186	0.631325301204818\\
187	0.635341365461846\\
188	0.639357429718874\\
189	0.643373493975902\\
190	0.64738955823293\\
};

\end{axis}
\end{tikzpicture}